\documentclass[%
 reprint,
superscriptaddress,
 amsmath,amssymb,
 aps
]{revtex4-2}
\pdfoutput=1

\usepackage{graphicx}% Include figure files
\usepackage{dcolumn}% Align table columns on decimal point
\usepackage{bm}% bold math
\usepackage{amsmath,mathrsfs,amssymb,dsfont}
\usepackage{xcolor}
\usepackage{import}
\usepackage{enumitem}
\usepackage[normalem]{ulem}
\usepackage{soul}
\usepackage{pdfpages}

\newcommand{\ad}{\mathrm{ad}}
\newcommand{\pert}{\mathrm{per}}

\newcommand{\z}{\hat{\mathbf{z}}}
\newcommand{\R}[1]{\mathbb{R}^{#1}}
\newcommand{\xx}{\mathbf{x}}
\newcommand{\bb}{\mathbf{b}}
\newcommand{\maxx}{\mathrm{max}}
\newcommand{\pauli}{\boldsymbol{\sigma}}
\newcommand{\sx}{\sigma_x}
\newcommand{\sy}{\sigma_y}
\newcommand{\sz}{\sigma_z}
\newcommand{\id}{\mathds{1}}
\newcommand{\stateI}{\vert \psi_0 \rangle}
\newcommand{\stateId}{\langle \psi_0 \vert}
\newcommand{\stateT}{\vert \psi_T \rangle}
\newcommand{\stateTd}{\langle \psi_T \vert}

\newcommand{\spinup}{\left\vert \uparrow \right\rangle}
\newcommand{\spindown}{\left\vert \downarrow \right\rangle}
\newcommand{\spinupd}{\left\langle \uparrow \right\vert}
\newcommand{\spindownd}{\left\langle \downarrow \right\vert}
\newcommand{\abs}[1]{\vert #1 \vert}
\newcommand{\dyson}[2]{\mathcal{D}_{#1}(#2)}

\makeatletter
\AtBeginDocument{\let\LS@rot\@undefined}
\makeatother

\begin{document}

\preprint{APS/123-QED}

\title{Numerical Engineering of Robust Adiabatic Operations}

\author{Sahand Tabatabaei}
\affiliation{Department of Physics, University of Waterloo, Waterloo, ON, Canada, N2L3G1}
\affiliation{Institute for Quantum Computing, University of Waterloo, Waterloo, ON, Canada, N2L3G1}

\author{Holger Haas}
\affiliation{Department of Physics, University of Waterloo, Waterloo, ON, Canada, N2L3G1}
\affiliation{Institute for Quantum Computing, University of Waterloo, Waterloo, ON, Canada, N2L3G1}

\author{William Rose}
\affiliation{Department of Physics, University of Illinois at Urbana-Champaign, Urbana, Illinois 61801, USA}

\author{Ben Yager}
\affiliation{Department of Physics, University of Waterloo, Waterloo, ON, Canada, N2L3G1}
\affiliation{Institute for Quantum Computing, University of Waterloo, Waterloo, ON, Canada, N2L3G1}

\author{Mich\`ele Piscitelli}
\affiliation{Department of Physics, University of Waterloo, Waterloo, ON, Canada, N2L3G1}
\affiliation{Institute for Quantum Computing, University of Waterloo, Waterloo, ON, Canada, N2L3G1}

\author{Pardis Sahafi}
\affiliation{Department of Physics, University of Waterloo, Waterloo, ON, Canada, N2L3G1}
\affiliation{Institute for Quantum Computing, University of Waterloo, Waterloo, ON, Canada, N2L3G1}

\author{Andrew Jordan}
\affiliation{Department of Physics, University of Waterloo, Waterloo, ON, Canada, N2L3G1}
\affiliation{Institute for Quantum Computing, University of Waterloo, Waterloo, ON, Canada, N2L3G1}

\author{Philip J. Poole}
\affiliation{National Research Council of Canada, Ottawa, Ontario, Canada, K1A 0R6}

\author{Dan Dalacu}
\affiliation{National Research Council of Canada, Ottawa, Ontario, Canada, K1A 0R6}

\author{Raffi Budakian}
\email{rbudakian@uwaterloo.ca}

\affiliation{Department of Physics, University of Waterloo, Waterloo, ON, Canada, N2L3G1}
\affiliation{Institute for Quantum Computing, University of Waterloo, Waterloo, ON, Canada, N2L3G1}
\affiliation{Canadian Institute for Advanced Research, Toronto, ON, Canada, M5G1Z8}

\date{\today}% It is always \today, today,

%%% Abstract %%%
\begin{abstract}
Adiabatic operations are powerful tools for robust quantum control in numerous fields of physics, chemistry and quantum information science.
The inherent robustness due to adiabaticity can, however, be impaired in applications requiring short evolution times.
We present a single versatile gradient-based optimization protocol that combines adiabatic control with effective Hamiltonian engineering in order to design adiabatic operations tailored to the specific imperfections and resources of an experimental setup.
The practicality of the protocol is demonstrated by engineering a fast, 2.3 Rabi cycle-long adiabatic inversion pulse for magnetic resonance with built-in robustness to Rabi field inhomogeneities and resonance offsets. 
The performance and robustness of the pulse is validated in a nanoscale force-detected magnetic resonance experiment on a solid-state sample, indicating an ensemble-averaged inversion accuracy of $99.997\%$.
We further showcase the utility of our protocol by providing examples of adiabatic pulses robust to spin-spin interactions, parameter-selective operations and operations connecting arbitrary states, each motivated by experiments.
\end{abstract}

\maketitle

%%% Main Body %%%
%%% Introduction %%%
\section{Introduction} \label{sec:introduction}
Since its inception by Born and Fock \cite{ref:BornFock}, the concept of adiabatic evolution has had a profound impact on quantum science and technology.
The adiabatic theorem establishes that under any sufficiently slow excitation, energy eigenstates of a quantum system remain eigenstates at all times \cite{ref:kato}.
The theorem has lent itself as an indispensable design principle for engineering quantum control sequences \cite{ref:abragam-1983, ref:Return, ref:SuperadiabaticNMR-2008, ref:DRAG-2009, ref:Gambetta-DRAG-extension-2011, ref:Guerin-2011, ref:Martinis-2014, ref:Chasseur-2015, ref:Superadiabatic-2016, ref:Petrosyan-2017, Vitanov-2017, ref:Quiroz-2019, ref:Khazali-2020} which, especially in the case of state-to-state transfers, exhibit remarkable robustness to a variety of experimental imperfections such as unitary control errors and decoherence \cite{ref:Robustness, ref:RobustnessDecoherence1, ref:RobustnessDecoherence2}.
Control sequences based on the principle of adiabaticity are widely used for quantum computation \cite{ref:AQCequivalence, Tosi-2017} as well as in quantum sensing, spectroscopy and imaging \cite{Grinolds-2013, Aiello-2013, Rendler-2017} with applications ranging from the early population transfers in magnetic resonance (MR) \cite{ref:abragam-1983,Slichter-2013} to recent implementations of quantum gates \cite{Kandala-2019, Arute-2019}.

True adiabaticity only occurs in the infinite duration limit, typically for closed quantum systems \cite{Venuti-2016}.
In real-world applications that demand fast operations, the robustness of the control sequence often degrades, because the evolution necessarily deviates from true adiabatic evolution.
Robustness may be further compromised due to unaccounted perturbation Hamiltonians and decoherence introduced by the coupling of the quantum system to external degrees of freedom.
Furthermore, shorter duration control sequences occupy a wider bandwidth, and are therefore more susceptible to distortions arising from bandwidth limitations of the control electronics.

The loss of robustness due to imperfect adiabaticity has evoked a need for better control engineering methods, which has led to various optimization schemes for adiabatic control
\cite{ref:Quiroz-2019,ref:AdiabaticOCT-2014,ref:AdiabaticOCTold-1996,ref:Return,ref:UnknownEnergyGap-2011}, as well as shortcuts to adiabaticity \cite{ref:Shortcuts-2019} such as superadiabatic operations \cite{ref:SuperadiabaticNMR-2008,ref:Superadiabatic-2016} and counter-diabatic driving \cite{ref:Demirplak1-2003,ref:Demirplak2-2005,ref:TransitionlessQuantumDriving-2009}.
In parallel to the development of more robust adiabatic controls, recent advances in Hamiltonian engineering techniques \cite{Haeberlen-1968, Green-2013, Soare-2014, Paz-Silva-2014, Choi-2020} provide a highly versatile means of designing high-fidelity quantum control sequences, tailored to specific applications.
Some of these approaches are based on numerical optimization methods that provide a systematic means of incorporating robustness to a variety of imperfections such as dominant decoherence sources \cite{Pasini-2009, Grace-2012, Soare-2014, ref:Haas_2019}.
Numerical control engineering \cite{Glaser-2015} may also be adapted to include ensemble effects such as inhomogeneities and parameter uncertainties \cite{Kobzar-2008, Borneman-2010, Li-2011, Goerz-2014, Bentley-2020}, as well as control resource limitations and distortions \cite{Motzoi-2011, Spindler-2012, Borneman-2012, Hincks-2015} in a particular experimental setup.

While several numerical schemes exist for either optimizing various adiabaticity metrics \cite{ref:Quiroz-2019,ref:AdiabaticOCT-2014,ref:AdiabaticOCTold-1996,ref:Return,ref:UnknownEnergyGap-2011}, or for minimizing the effect of perturbation Hamiltonians \cite{Pasini-2009, Grace-2012, Soare-2014, ref:Haas_2019}, a single versatile numerical optimization protocol that combines adiabatic control with effective Hamiltonian engineering would be highly valuable across quantum information science.
With this work, we present a unified protocol for engineering adiabatic operations that are robust to various perturbations, including certain decoherence pathways that may arise in cases where the system evolution deviates from perfect adiabaticity, in particular applications that require short control times.
Our work relies on extending the Van Loan formalism \cite{ref:Haas_2019} to allow for efficient optimization of adiabaticity and various perturbation terms on an equal footing.
The Van Loan formalism utilizes auxiliary block matrices \cite{ref:VL, ref:CARBONELL, ref:Goodwin, Machnes-2018, Goodwin-2020,nonHolonomic} for efficiently evaluating a broad family of integrals involving the time evolution operator, while also allowing for gradient-based optimization for quickly convergent control searches.

In this paper, we focus on adiabatic control engineering for two-level quantum systems, i.e. qubits, which we refer to as ``spins'' in the context of our MR experiments.
Our protocol can, however, be easily extended to multi-level systems in which the energy eigenstates are analytically known as a function of the controls. An example of this is given in Ref. \cite{ref:Supplements} for $s>1/2$ nuclear spins in MR.
In the following, we will first outline the adiabatic control problem and the details of the optimization scheme.
We then demonstrate the utility and practicality of the protocol by designing adiabatic passages for MR and conducting an experimental study of a fast, $2.3$ Rabi cycle-long adiabatic inversion pulse engineered to be robust to both large Rabi field variations and resonance offsets.
The measurements are carried out on an interacting solid-state nanoensemble of nuclear spins using a force-detected MR setup with an intrinsically high degree of Rabi field variations.
The ensemble of spins experiences a continuous range of Rabi fields that varies in amplitude by $2\times$.
We find that the $z-$magnetization decays to $e^{-1}$ of its initial value after $\sim 34,000$ successive applications of the inversion pulse, corresponding to an ensemble-averaged single-pulse inversion accuracy of 99.997\%. We further characterize the pulse performance as a function of Rabi field and resonance offset, and compare these measurements with the ideal calculated performance for this pulse.

The potential utility of our protocol is further demonstrated with three self-contained examples in the appendices.
We describe an adiabatic inversion pulse designed to be robust against strong dipole-dipole interactions in a dense network of electron spins. We provide numerical results that demonstrate that by including effective Hamiltonian terms that minimize dipolar interactions into the optimization protocol, the pulse fidelity can be significantly improved. We further show that our protocol can be used to design adiabatic pulses that are selective to a particular set of Hamiltonian parameters -- an important feature for sensing and spectroscopy applications \cite{Degen-2017, Barry-2020}. We conclude by discussing the application of our protocol to connect arbitrary states.

%%% Adiabatic Control Problem %%%
\section{Adiabatic Control Problem}\label{sec:setup}
Consider the dynamics of a spin over an interval $[0, T]$ governed by a Hamiltonian $H$ which depends on a set of \textit{control parameters} $\xx = (x_1,...,x_N) \in \R{N}$.
The control parameters indicate the experimenter's ability to control the spin, and here, we regard them as a particular parametrization of an electromagnetic waveform that couples to the spin.
The $\xx$ vector could, for example, be a set of the waveform's piecewise constant amplitudes (e.g. to be implemented on an arbitrary waveform generator).
Any such Hamiltonian can be expanded as
\begin{equation}
    H\big(\bb(\xx,t)\big) = - \bb(\xx,t)\cdot \pauli/2, \label{eq:GeneralHamiltonian}
\end{equation}
where $\pauli \equiv (\sx,\sy,\sz)$ denotes a vector of Pauli matrices, and the function $\bb : \R{N}\times[0,T] \rightarrow \R{3}$ is assumed to be differentiable, as required for gradient-based optimization.
The exact form of $\bb(\xx,t)$ varies for different experimental settings; here, we will assume that this functional relationship is known.
In Section~\ref{sec:adiabaticMR} we provide a concrete example of a $\bb(\xx,t)$ when discussing adiabatic MR pulses.
The dynamics of the spin is thus completely determined by the trajectory of $\bb(\xx,t)$, which borrowing from MR terminology, we refer to as the \textit{effective field} throughout this paper.
The eigenvalues of Eq.(\ref{eq:GeneralHamiltonian}) are given by {$\pm \abs{\bb(\xx,t)}\big /2$,} with the corresponding eigenstates pointing along the $\pm \bb(\xx,t)$ direction on the Bloch sphere.

The basic form of the adiabatic control problem is to find a set of control parameters $\bar{\xx}$ such that given some initial state $\stateI$ and target state $\stateT$, 
\begin{enumerate}[label=\Alph*.]
    \item $\stateI$ evolves to $\stateT$ at $t=T$, up to a phase. \label{Statement1}
    \item {The} state follows a specific instantaneous eigenstate of {$H\big(\bb(\bar{\xx},t)\big)$}, as closely as possible, for all $t\in [0,T]$. \label{Statement2}
\end{enumerate}
Condition \ref{Statement2} implies that $\bb(\bar{\xx},0)$ and $\bb(\bar{\xx},T)$ are constrained such that they point along the $\vert \psi_0 \rangle$ and $\vert \psi_T \rangle$ directions on the Bloch sphere, respectively.

In addition to the above criteria for adiabatic evolution, we generally require the operation to be robust to various experimental imperfections and uncertainties.
For example, protection against a single-body perturbation Hamiltonian $\delta H(t)$ amounts to minimizing the variation of the final state $\vert \delta \psi(\xx) \rangle$ due to the perturbation, which can be computed using the Dyson series \cite{ref:Dyson}:
\begin{equation}
    \vert \delta \psi(\xx) \rangle = \sum_{n=1}^\infty \mathcal{D}_U(\underbrace{\delta H,..., \delta H}_{n \text{ times}};T) \stateI, \label{eq:StateVariation}
\end{equation}
where we use a shorthand
\begin{eqnarray}
    &&\dyson{\mathcal{U}}{A_1,...,A_n; t} \equiv (-i)^n\mathcal{U}(t) \nonumber\\  && \ \times \int_0^t dt_1 \dots \int_0^{t_{n-1}} dt_n \prod_{k = 1}^n \Big(\mathcal{U}^{-1}(t_k) A_k(t_k) \mathcal{U}(t_k) \Big), \label{eq:DysonTerms}
\end{eqnarray}
for the various \textit{Dyson terms}, along with $U(\xx,t) = \mathrm{Texp}\big[-i\int_0^t dt' H \big(\bb(\xx,t')\big)\big]$ for the unperturbed propagator, with $\mathrm{Texp}$ denoting the time-ordered exponential.
In practice, one truncates the infinite series in Eq.(\ref{eq:StateVariation}) at some finite order and minimizes the final state correction up to that order.

In this paper, we will restrict our attention to single-body perturbations at first order in the Dyson series, which is commonly referred to as the zeroth order average Hamiltonian in MR literature \cite{ref:AHT}.
Nevertheless, the Van Loan formalism \cite{ref:Haas_2019} employed here allows for the inclusion of many-body perturbations up to arbitrary order.

When assessing the suitability of a particular set of control parameters $\xx$ we use three metrics that quantify the final state fidelity, adiabaticity of the control trajectory and its sensitivity to perturbations.
We quantify the final state fidelity by calculating the overlap function $\varphi_0(\xx) \equiv \abs{\stateTd U(\xx,T) \stateI}^2$. We evaluate adiabaticity using the time-averaged overlap between $U(\xx,t) \stateI$ and the instantaneous energy eigenstate $\vert E_\pm (\bb)\rangle$,
\begin{eqnarray}
    \varphi_\ad(\xx) \equiv \frac{1}{T} \int_0^T dt \Big\vert \big\langle E_\pm\big(\bb(\xx,t)\big) \big\vert U(\xx,t) \big\vert \psi_0 \big\rangle \Big\vert^2\nonumber\\
     = \frac{1}{T} \stateId U^\dagger(\xx,T) \dyson{U}{iP;T} \stateI, \label{eq:AdiabaticityDyson}
\end{eqnarray}
where $P(\bb) = (\id \pm \bb \cdot \pauli/ \abs{\bb})/2$ is the projection operator onto the $\pm\abs{\bb}$ eigenspace of $H(\bb)$. The $\pm$ sign here is chosen such that $\stateI$ coincides with the corresponding eigenstate at $t=0$.
Here, we have assumed that $\abs{\bb} \neq 0$ for all $t$.
In general, $0 \leq \varphi_0(\xx)\leq 1$ and $0\leq \varphi_\ad(\xx) \leq 1$, whereas $\varphi_0(\xx)= \varphi_\ad(\xx) = 1$ if and only if conditions \ref{Statement1} and \ref{Statement2} are perfectly satisfied.
Lastly, we quantify robustness to perturbation Hamiltonians using Eq.(\ref{eq:StateVariation}) by defining $\varphi_\pert(\xx) \equiv 1 - \Vert \delta\psi(\xx)\Vert^2/\mathcal{N}^2$, or to first order in $\delta H$:
\begin{equation}
    \varphi_\pert(\xx) \equiv 1 - \frac{1}{\mathcal{N}^2} \stateId \dyson{U}{\delta H,T}^\dagger \dyson{U}{\delta H,T} \stateI, \label{eq:PerturbationMetric}
\end{equation}
with $\mathcal{N}=\int_0^T dt \ \Vert \delta H(t) \Vert_\text{op}$ being a normalization ensuring $0\leq\varphi_\pert(\xx) \leq 1$, given $\big\Vert \delta H(t) \big\Vert_\mathrm{op} \equiv \mathrm{sup}_{\vert v \rangle \neq 0}\Big(
\big\Vert\delta H(t) \vert v \rangle \big\Vert/\Vert v \Vert\Big)$.

Another important consideration for practical control design are ensemble effects that manifest themselves as variations in the parameters of the system, e.g., non-uniform static/control fields, parameter uncertainties and ensembles in time, i.e. different conditions in distinct experimental realizations.
These can be dealt with by considering a collection of quantum systems $\Lambda$, with each member $\lambda \in \Lambda$ evolving under a different value for the parameters in question.
Robustness to such ensemble effects translates to finding the optimum control parameters that result in an evolution satisfying conditions \ref{Statement1} and \ref{Statement2} for all $\lambda \in \Lambda$.

%%% Control Engineering  %%%
\section{Control Engineering} \label{sec:theory}
For simplicity, we first discuss our adiabatic control protocol for a single spin, without accounting for inhomogeneities or uncertainties in the system.
The full protocol is then discussed accordingly.
A solution $\bar{\xx}$ to the single-spin adiabatic control problem can be found by setting up and numerically maximizing a combined target function:
\begin{equation}
    \varphi \equiv p_0\varphi_0+ p_\ad \varphi_\ad+ p_\pert\varphi_\pert, \label{eq:totalTarget}
\end{equation}
where the relative weights $p_0,\ p_\ad, \ p_\pert$ are non-negative and add up to $1$.
To efficiently compute the integral terms arising in $\varphi_0$, $\varphi_\ad$ and $\varphi_\pert$ as well as the gradients of the integral terms with respect to $\xx$, we utilize the Van Loan relations for the time-ordered exponential of upper-triangular block matrices \cite{ref:VL,ref:Haas_2019,ref:CARBONELL,ref:Goodwin}.
We use a \textit{Van Loan generator}
\begin{equation}
     L(\bb,t) \equiv -i\begin{bmatrix}
    H(\bb) & iP(\bb) & 0\\[7pt]
    0 & H(\bb) & \delta H(t)\\[7pt]
    0 & 0 & H(\bb)
\end{bmatrix},\label{eq:VLgenerator}
\end{equation}
to construct the \textit{Van Loan propagator},
\begin{eqnarray}
     V_t[\bb] &&\equiv \mathrm{Texp}\Big[\int_0^t dt' L\big(\bb(\xx,t'),t'\big)\Big]\nonumber\\[5pt]
    &&=\begin{bmatrix}
        U(\xx,t) & \mathcal{D}_U(iP;t) & \mathcal{D}_U\big(iP,\delta H; t\big)\\[10pt]
        0 & U(\xx,t) & \mathcal{D}_U(\delta H;t)\\[10pt]
        0 & 0 & U(\xx,t)
    \end{bmatrix} \label{eq:VLpropagator}
\end{eqnarray}
\cite{ref:Haas_2019}.
It is clear from Eq.(\ref{eq:VLpropagator}) that the Van Loan formalism allows for the simultaneous evaluation of all quantities needed for the computation of Eq.(\ref{eq:totalTarget}), which are contained in the different blocks of $V_T[\bb]$.

\begin{figure}[h]
    \includegraphics[width = \columnwidth]{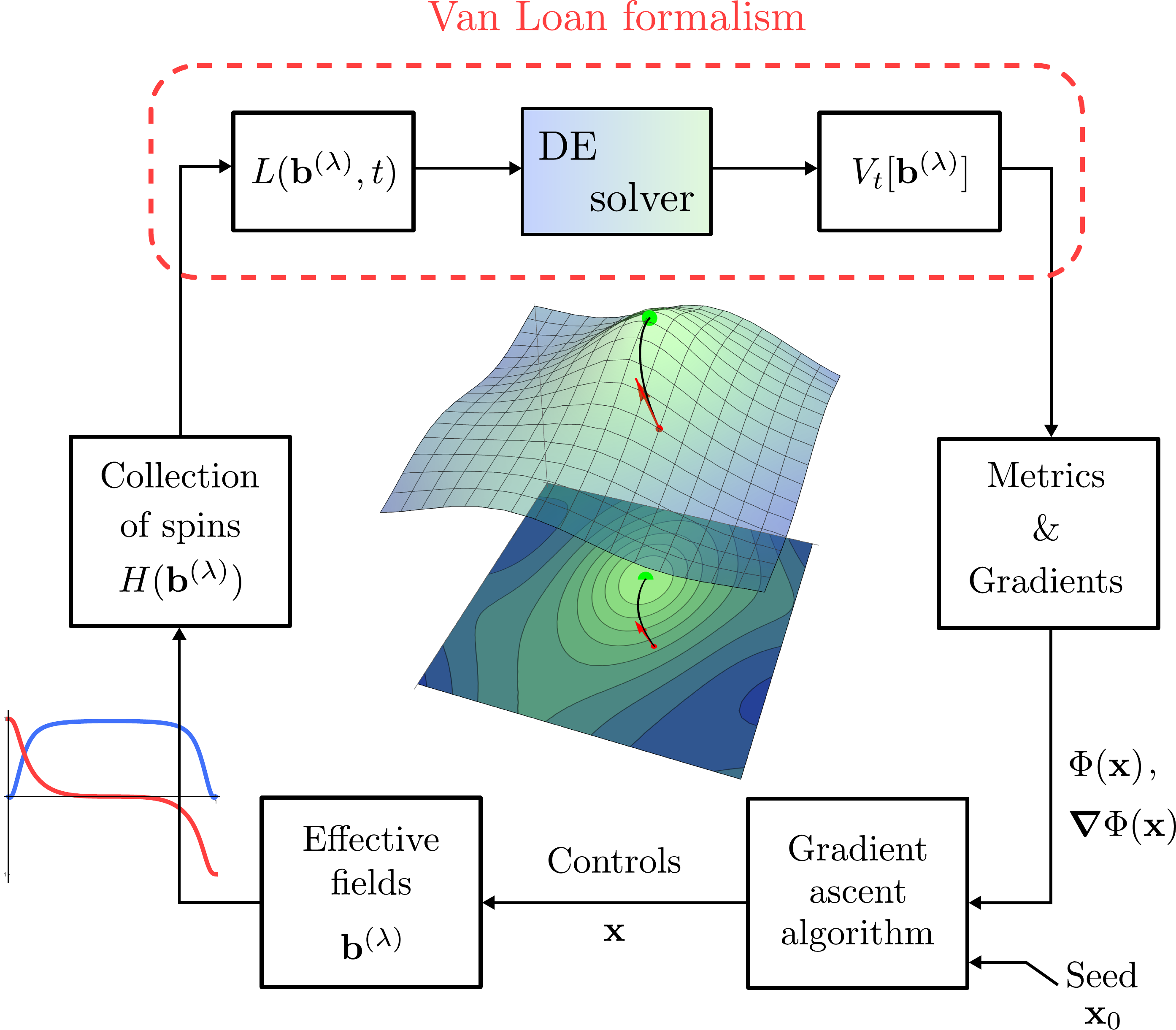}
    \caption{
        \label{fig:blkDiagram} 
        General overview of the adiabatic control protocol.
        In each iteration of the optimization algorithm, the control parameters $\xx$ are used to calculate the effective field trajectory, and subsequently the Hamiltonian, for each member of the optimization {set}.
        We then utilize the Van Loan auxiliary matrix formalism to evaluate the various control performance metrics, along with their gradients, and provide them to the gradient ascent optimizer.
    }
\end{figure}

A schematic of our full adiabatic control engineering protocol is depicted in Fig. \ref{fig:blkDiagram}. We consider a finite collection of spins $\Gamma \subseteq \Lambda$, called the \textit{optimization set}, with size $\abs{\Gamma}$, that serves as a representative subset of spins in the experiment, which experience the ensemble effects we wish to address.
The protocol starts with a seed $\xx_0$ for the control parameters that is chosen from a pseudorandom distribution, and is then used to calculate the Van Loan generator $L\big(\bb^{(\lambda)}(\xx_0,t)\big)$ for each member $\lambda \in \Gamma$. A numerical differential equation (DE) solver, such as a Runge-Kutta algorithm \cite{ref:RungeKutta}, is then used to calculate the Van Loan propagators $\{V_t[\bb^{(\lambda)}]\}$, which are then used to
evaluate the single-member target functions $\varphi^{(\lambda)}(\xx)$, given by Eq.(\ref{eq:totalTarget}), for every $\lambda \in \Gamma$. The set of target functions $\{\varphi^{(\lambda)}(\mathbf{x})\}$ are combined to form a total target function $\Phi \equiv \sum_{\lambda\in \Gamma} w^{(\lambda)} \varphi^{(\lambda)}$, to be maximized by an optimization algorithm.
The relative weights $w^{(\lambda)}$ prioritize different members of the optimization set, and satisfy $0 \leq w^{(\lambda)} \leq 1$, with $\sum_{\lambda \in \Gamma } w^{(\lambda)} = 1$. 

For efficient control engineering we also need to evaluate the target function gradient $\boldsymbol{\nabla} \Phi(\xx)$, which together with $\Phi(\xx)$, is supplied to a gradient ascent optimizer to search the control landscape for the optimum adiabatic operation.
Since $\lbrace \varphi^{(\lambda)} \rbrace$ are simple functions of $\lbrace V_T[\bb^{(\lambda)}] \rbrace$, their partial derivatives with respect to the control parameters can be evaluated through the chain rule by noticing that
\begin{equation}
   \frac{\partial V_T[\bb]}{\partial \xx} = \int_0^T dt \ \frac{\delta V_T[\bb]}{\delta \bb(\xx,t)} \cdot \frac{\partial \bb(\xx,t)}{\partial \xx} , \label{eq:ChainRuleIntegral}
\end{equation}
where $\delta/\delta \bb \equiv (\delta/\delta b_x,\delta /\delta b_y,\delta /\delta b_z)$,
\begin{equation}
    \frac{\delta V_T[\bb]}{\delta b_\alpha(\xx,t)} = V_T[\bb] \ V_t^{-1}[\bb] \  \frac{\partial L (\bb,t)}{\partial b_\alpha} \ V_t[\bb]
    \label{eq:VLpropagatorFunctionalDeriv}
\end{equation}
and $\partial \bb(\xx,t)/\partial \xx$ is the $3\times N$ Jacobian matrix of the effective field. The matrix inverses appearing in the expression for the functional derivatives $\delta V_T[\bb]/\delta b_\alpha(\xx,t)$, $\alpha \in \{x,y,z\}$, can be evaluated analytically \cite{ref:matInv}.
Eq.(\ref{eq:ChainRuleIntegral}) and Eq.(\ref{eq:VLpropagatorFunctionalDeriv}), along with the assumption that the functional form of $\bb(\xx,t)$ is known, imply that $\boldsymbol{\nabla} \Phi(\xx)$ can be computed using the values of $V_t[\bb]$, which are already available from the output of the DE solver.
In practice, we approximate the integral in Eq.(\ref{eq:ChainRuleIntegral}) by a finite sum for computational speedup. 

%%% Adiabatic Pulse Design for Magnetic Resonance %%%
\section{Adiabatic Pulse Design for Magnetic Resonance}\label{sec:adiabaticMR}
We now discuss the application of the proposed control protocol to the design of adiabatic passages for MR.
Consider a spin $1/2$ particle with gyromagnetic ratio $\gamma$ in a static magnetic field $\mathbf{B}_0 = B_0 \z$, and an orthogonal radio-frequency (RF) field $\mathbf{B}_\text{RF}(t) = 2 B_{1I}(t) \cos \phi(t) + 2 B_{1Q}(t) \sin \phi(t)$, where $B_{1I}(t)$ and $B_{1Q}(t)$ are the in-phase and quadrature envelope functions, and the instantaneous angular frequency is $\dot{\phi}(t)$.
Measuring energy in units of angular frequency, the Hamiltonian of the spin in a reference frame rotating around $\z$ at the instantaneous RF frequency is known to be \cite{ref:Return}
\begin{equation}
    H(t) = - \omega_{1I}(t) \frac{\sx}{2} - \omega_{1Q}(t) \frac{\sy}{2} - \Delta \omega(t) \frac{\sz}{2}, \label{eq:MRHamiltonian}
\end{equation}
where $\omega_{1I}(t) \equiv \gamma B_{1I}(t)$, $\omega_{1Q}(t) \equiv \gamma B_{1Q}(t)$ and $\Delta \omega(t) \equiv \gamma B_0- \dot{\phi}(t)$ is the instantaneous resonance offset.
Comparing Eq.(\ref{eq:GeneralHamiltonian}) and Eq.(\ref{eq:MRHamiltonian}) reveals that $\bb(t) \equiv \big(\omega_{1I}(t),\omega_{1Q}(t),\Delta \omega(t)\big)$ is the effective field in this problem.
The adiabatic control problem here amounts to finding the optimal envelope and resonance offset waveforms that maximize the target function discussed in Section \ref{sec:theory}.
We will now focus on engineering adiabatic full passages (AFPs), that adiabatically evolve spins between the states $\spinup$ and $\spindown$.

In order to formulate the protocol in terms of a finite set of optimization parameters $\xx \in \R{N}$, we need to parametrize the three waveforms that represent $\bb(t)$ using an ansatz suitable for the desired operation.
An appropriate choice of parametrization should accommodate the initial and target states, and be flexible enough to handle the imperfections in question, while involving a minimal number of unknowns in the optimization, and hence more efficient pulse searches.
Accordingly, we parametrize the waveforms as $b_x(\xx,t) = \omega_{1\maxx} \tanh [a_x(\xx,t)]$, $b_y(\xx,t) = 0$ and $b_z(\xx,t) = \Delta \omega_\maxx \tanh [a_z(\xx,t)]$, where
\begin{equation}
    \begin{cases}
        \displaystyle a_x(\xx,t) \equiv  \sum_{n=1}^{N/2} x_n \bigg[1- \big(1-2 \frac{t}{T}\big)^{2n}\bigg]\\[15pt]
        \displaystyle a_z(\xx,t) \equiv  \sum_{n=N/2+1}^{N} x_n \Big(1-2 \frac{t}{T}\Big)^{2(n-N/2)-1}
    \end{cases}, \label{eq:Ansatz}
\end{equation}
are even and odd polynomials around $t=T/2$, respectively.
The parameters $\omega_{1\maxx}$ and $\Delta \omega_\maxx$ are the maximum Rabi strength and resonance offset realizable in the experimental setup.
The hyperbolic tangent functions act as soft clipping functions that restrict the waveforms according to the chosen maximum values, while $\Delta \omega_\maxx$ serves as a handle on the bandwidth of the frequency-modulated pulse, which is roughly on the order of $2 \Delta \omega_\maxx$.
We briefly discuss a more general ansatz for operations connecting arbitrary states in App. \ref{app:arbitraryState}.

We now provide an example AFP engineered for a spin ensemble experiencing a broad range of maximum Rabi strengths $\omega_{1\maxx} \in [\Omega_1, 2 \Omega_1]$, for some $\Omega_1$.
Analytically derived AFPs with similar properties \cite{ref:Return} have been used as part of a detection protocol in recent force-detected nanoscale magnetic resonance experiments \cite{ref:PRX}.
We normalize frequency and time variables in units of the smallest Rabi frequency $\Omega_1$, and its associated Rabi cycle $2\pi/\Omega_1$, respectively.
The normalized duration of the pulse is set to $T \Omega_1/(2 \pi) = 2.3$.
Optimization is done on a 5-element optimization set, with the maximum resonance offset $\Delta \omega_\maxx/\Omega_1 = 5$, as an approximate bandwidth constraint, and an additional perturbation metric [$\delta H(t) = \sigma_z$] for robustness against Larmor inhomogeneities.
$\abs{\Gamma} = 5$ is chosen empirically by monitoring the Rabi-frequency-dependence of the various pulse metrics in trial optimizations, adding more elements if the pulse performance is not satisfactory over the whole $[\Omega_1,2 \Omega_1]$ Rabi range.
The waveforms are parametrized with $N=50$ coefficients, using the polynomial ansatz of Eq.(\ref{eq:Ansatz}).
The relative weights in Eq.(\ref{eq:totalTarget}) are set to $p_0^{(\lambda)} =  0.2$, $p_\ad^{(\lambda)} = 0.6$ and $p_\pert^{(\lambda)} = 0.2$ for all $\lambda \in \Gamma$. The values of $p_0$, $p_\ad$ and $p_\pert$ are also chosen empirically to ensure the simultaneous convergence of each individual metric $\varphi_0$, $\varphi_\ad$ and $\varphi_\pert$ when optimizing $\varphi$ \cite{ref:Haas_2019}, while slightly emphasizing $p_\ad^{(\lambda)}$ to increase the control robustness arising from adiabaticity.
The weights for optimization-set members are chosen as $w^{(\lambda)} = 1/\vert \Gamma \vert = 1/5$.
For the gradient ascent optimizer and DE solver, we use the FindMaximum function and the explicit Runge-Kutta algorithm of the ParametricNDSolve function in Mathematica \cite{ref:Mathematica}, respectively.
Initialization of the optimizer is done by drawing a seed from a uniform pseudorandom distribution on $[-1, 1]^{50}$, resetting the optimizer with a new seed if the target function value does not exceed 0.99 after 50 steps.
The computations for different optimization-set members are parallelized on a multi-core processor for additional speedup.

\begin{figure}[h]
    \includegraphics[width = \columnwidth]{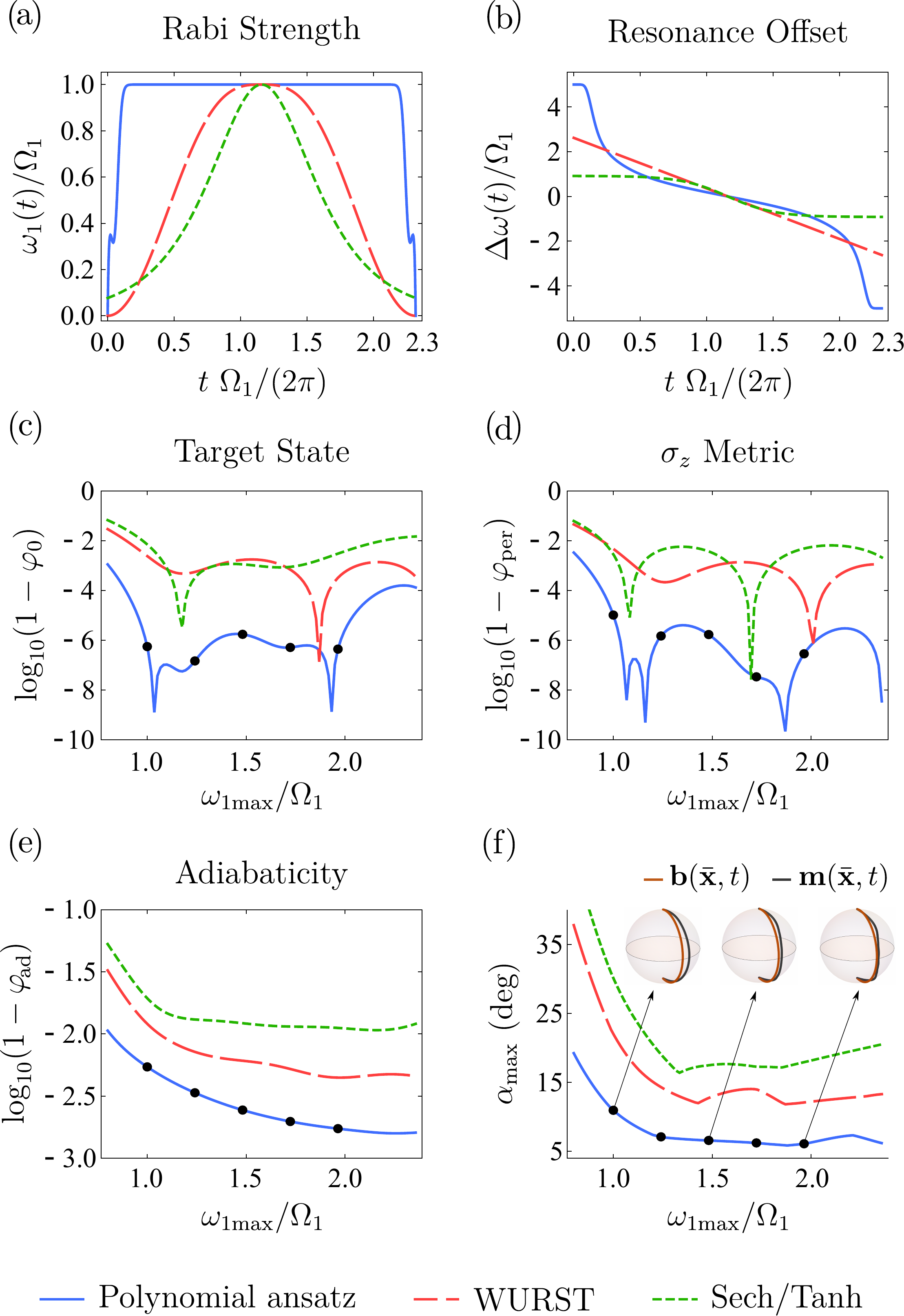}
    \caption{(a) Amplitude and (b) resonance offset waveforms of the numerically optimized fast AFP using the polynomial ansatz, along with the WURST and Sech/Tanh reference pulses.
    (c) Target state, (d) $\sigma_z$ perturbation and (e) adiabaticity metrics for the various AFPs as a function of maximum Rabi frequency, measured as their (logarithmic) deviation from unity.
    (f) The maximum angle between the effective field and the magnetization vectors throughout the evolution, along with Bloch sphere trajectories at the start, middle and end of the design Rabi range.
    The black dots in (c-f) indicate the five Rabi frequency values and control metrics for the optimization-set elements $\lbrace \lambda \rbrace$. \label{fig:AFP16} 
    }
\end{figure}

To benchmark the performance of the pulse, we use the same target function and 5-element optimization set, with identical amplitude and resonance offset constraints to optimize two reference AFPs of the same duration, using two standardized waveforms in the literature: the WURST \cite{ref:WURST1,ref:WURST2} pulse, used for fast, broadband spin inversions, and the Sech/Tanh \cite{ref:Sech1,ref:Sech2} pulse, which is known for its insensitivity to RF field variations above its cutoff Rabi frequency \cite{ref:Return} (see Ref. \cite{ref:Supplements} for the optimization details).
The optimized polynomial AFP, along with the two reference pulses are depicted in Fig. \ref{fig:AFP16}(a,b).
In the same figure, we also examine the Rabi-dependence of various control metrics.
The (logarithmic) infidelities associated with the target state and $\sigma_z$ perturbation metrics are plotted as a function of maximum Rabi strength in Fig. \ref{fig:AFP16}(c-d); indicating that the polynomial AFP infidelities are approximately two orders of magnitude lower than the ones for the two reference pulses, over the relevant Rabi range.
The polynomial AFP also exhibits a significantly higher degree of adiabaticity than the WURST and Sech/Tanh pulses, as can be seen in Fig. \ref{fig:AFP16}(e).
Fig. \ref{fig:AFP16}(f) shows another, more intuitive measure of adiabaticity -- the maximum angle between the magnetization and effective field vectors
\begin{equation} 
    \alpha_\maxx = \underset{t \in [0,T]}{\maxx} \cos^{-1} \bigg(\frac{\bb(\bar{\xx},t) \cdot \mathbf{m}(\bar{\xx},t)}{\big\vert \bb(\bar{\xx},t) \big\vert \big\vert \mathbf{m}(\bar{\xx},t) \big\vert}\bigg),
\end{equation}
where $\bar{\xx}$ is the set of optimal control parameters, and $\mathbf{m}(\bar{\xx},t) = \spinupd U^\dagger(\bar{\xx},t) \pauli U(\bar{\xx},t) \spinup/2$ is the magnetization.
In the design $\omega_{1\maxx}$ range, the plot indicates an ${\alpha_\maxx \leq 11^\circ}$ for the polynomial AFP, whereas the WURST and Sech/Tanh pulses reach ${\alpha_\maxx=22^\circ}$ and ${\alpha_\maxx=30^\circ}$, respectively.
Note that even though the polynomial AFP shows far better adiabaticity than the reference pulses, it still clearly deviates from perfect adiabatic evolution due to its very short duration.
Nevertheless, as will be shown experimentally in Section \ref{sec:experimental}, the pulse still possesses the desired robustness in a practical setting.

Motivated by quantum sensing experiments on nitrogen vacancy centers \cite{ref:Genov-2020}, we provide another example that demonstrates the utility of our perturbative treatment in App. \ref{app:el-AFPS}, where we engineer AFPs for densely-packed electrons, and specifically minimize the perturbations due to spin-spin dipolar interactions. %add citations
Numerically comparing the results with an AFP optimized without the perturbation metric shows an order of magnitude improvement in the target state infidelity for a system of 7 interacting electron spins.

The applicability of our adiabatic control protocol is not only limited to finding fast robust adiabatic operations, but also can be used to engineer control sequences that are selective in a particular Hamiltonian parameter; a common demand in sensing, imaging and spectroscopy applications \cite{Degen-2017, Barry-2020}.
For instance, adiabatic inversions that are selective to Larmor frequency are used for the spatially localized excitation of spins in some implementations of nanometer-scale magnetic resonance imaging (MRI) measurements \cite{ref:mrfm, ref:Degen,ref:nanoMRI}.
Engineering adiabatic pulses that exhibit well-defined, narrow inversion bands in Larmor frequency, with sharp band edges is an essential ingredient for achieving the highest resolution imaging in these nano-MRI approaches.
In App. \ref{app:slice-selective-AFPS} we demonstrate the application of our protocol to engineer adiabatic inversions that act over some chosen Larmor frequency range, with a specifically tailored inversion profile.

%%% Experimental Verification %%%
\section{Experimental Verification}\label{sec:experimental}
In this section, we present an experimental study of the 2.3 Rabi cycle AFP discussed in Section \ref{sec:adiabaticMR}, using the force-detected nano-MRI setup discussed in \cite{ref:PRX}.
The measurements are made on ${}^{31}$P nuclear spins in an indium phosphide (InP) nanowire sample \cite{Dan}, grown with a Wurtzite structure, inside a static field $B_0 = 3$~T applied along the growth axis ($z-$direction) at 6~K.
At this field, the Larmor frequency of ${}^{31}$P-spins, with gyromagnetic ratio $\gamma/(2 \pi) = 17.235$~MHz/T, equals $\omega_0/(2 \pi) = 51.8$~MHz.
The sample is a solid-state spin system with both homonuclear (P$\leftrightarrow$P) and heteronuclear (P$\leftrightarrow$In) dipolar and J couplings \cite{Slichter-2013}.
We calculate the nearest-neighbor dipolar coefficients to be $132$~Hz for P$\leftrightarrow$P couplings, and $619$~Hz for P$\leftrightarrow$In couplings.
Based on previous measurements of InP with zinc-blende crystal structure \cite{ref:Pines,ref:Iijima_2003}, we expect the P$\leftrightarrow$In and P$\leftrightarrow$P J coupling coefficients to be less than, or on the order of 1 kHz and 20 Hz, respectively.
Consistent with these expectations, Ramsey and Hahn echo experiments on our sample indicate relaxation times of $T_2^*=70 \ \mu$s and $T_2 = 364 \ \mu$s, respectively.

\begin{figure*}
\includegraphics[width = \textwidth]{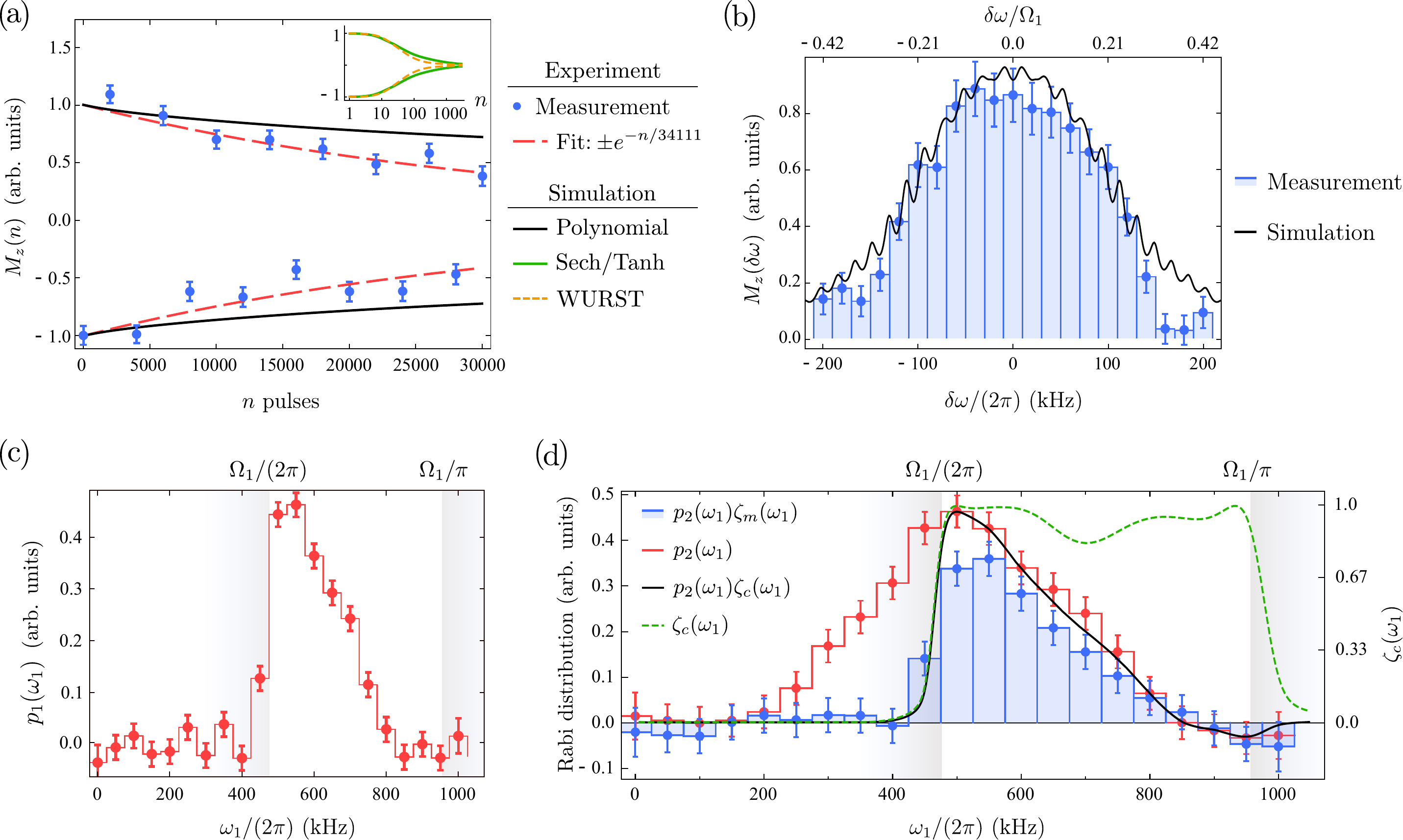}
\caption{Experimental study of the 2.3 Rabi cycle-long polynomial AFP.
(a) Spin signal after $n$ AFP applications, an exponential fit to the data and simulations of the optimized polynomial, WURST and Sech/Tanh pulses.
(b) Spin signal after 5000 AFPs, measured as a function of resonance offset.
The frequency resolution in the measurement is 20 kHz.
(c) Measured effective spin density $p_1(\omega_1)$ used in the fidelity and resonance offset measurements.
(d) Distributions $p_2(\omega_1) \zeta_m(\omega_1)$, $p_2(\omega_1)$ and $p_2(\omega_1) \zeta_c(\omega_1)$ (left axis), where $\zeta_c(\omega_1)$ is the calculated response function for a train of 5000 AFPs (right axis).
The frequency resolution of the Rabi frequency spectra shown in (c) and (d) is 50 kHz, with the shaded areas indicating Rabi frequencies outside the AFP design range.
All error bars correspond to one standard deviation.
\label{fig:Experimental}}
\end{figure*}

Here, we omit the details of the experimental setup and spin detection protocol, which are given in \cite{ref:PRX}. 
The $^{31}$P spins are controlled by RF magnetic fields near $\omega_0$ using a current-driven nanometer scale field source, which generates a highly non-uniform field profile.
The spins within our roughly $(100\text{-nm})^3$ detection volume experience a continuous range of Rabi frequencies from 172~kHz to  862~kHz.
Our measurements are done using the MAGGIC spin detection protocol \cite{ref:PRX}, which measures the integrated $z$-magnetization of the spin ensemble
\begin{equation}
    M_z \propto \int_0^\infty d \omega_1~  p(\omega_1) \text{Tr} \left[ \rho(\omega_1) \sigma_z \right] = \int_0^\infty d \omega_1~ p(\omega_1) \zeta (\omega_1), \label{eq:integrated-spin-signal}
\end{equation}
where $\rho(\omega_1)$ is the effective density operator of the spin at $\omega_1$, and $p(\omega_1)$ is an effective Rabi-frequency-dependent density of spins, which is determined by the device and sample geometry, as well as the parameters of the detection protocol \footnote{In contrast to conventional MR, we measure the average statistical correlation in the $z$ component of the spins, which dominates the thermal polarization for nanoscale spin ensembles. For the sake of interpreting our results, the measured signal can be thought of as the integrated local $z$-magnetization over an adjustable range of Rabi frequencies. See Ref.\cite{ref:PRX} for the details of the measurement procedure.}.
$\zeta(\omega_1)$ is the response function that characterizes the Rabi-frequency dependent performance of the sequence of AFPs under study.
The Rabi range in which $p(\omega_1)$ is non-zero can be tuned by adjusting the detection protocol, which allows us to measure only those spins that experience a specific range of Rabi frequencies.
To determine the Rabi frequency distribution of $M_z$, we use the Fourier encoding method described in \cite{ref:PRX}. The distributions $p_1(\omega_1)$ and $p_2(\omega_1)$ [Fig. \ref{fig:Experimental}(c,d)] used in the characterization of the 2.3 Rabi cycle AFP were measured separately.

We characterize a numerically engineered 2.3 Rabi cycle AFP that is 4.8~$\mu\text{s}$ long with a lower cutoff Rabi frequency set to $\Omega_1/(2\pi)=479$~kHz and the maximum resonance offset set to $\Delta\omega_{\text{max}} = 5\Omega_1/(2\pi) = 2.4$~MHz.
By Fourier transforming the pulse waveform, we determine the bandwidth of the pulse to be 5.4 MHz -- roughly equal to $2 \Delta \omega_\maxx$.
We test the AFP in three different aspects: overall fidelity, as well as robustness to resonance offsets and Rabi frequency inhomogeneities.
The $\sigma_z$ metric used in the optimization assists the pulse performance in the presence of small static-field inhomogeneities, chemical shifts and heteronuclear couplings.
Because of the relatively weak dipole-dipole interactions in InP, we did not include a perturbation metric for homonuclear dipolar interactions into the pulse optimization. Multi-spin simulations confirmed that dipolar interactions produced minimal degradation in the pulse fidelity.
While dipolar interactions are not a significant perturbation for the InP system, there are instances where dipolar interactions can significantly affect the pulse performance. As an example, in App. \ref{app:el-AFPS} we consider numerically adiabatic inversions for a dense network of electron spins.

To examine the fidelity, we confine the measured Rabi range $[\Omega_1,2 \Omega_1]$ to match the design Rabi range of the 2.3 Rabi cycle AFP.
The corresponding measured distribution $p_1(\omega_1)$ is shown in Fig. \ref{fig:Experimental}(c).
We apply $n$ consecutive AFPs and measure the decay in $M_z$ as a function of $n$, where $n\in[1,30016]$.
When performing the experiment, we insert $t_w = 52~\mu$s delays between the applied AFPs, which act as effective dephasing maps \cite{ref:NielsenChuang} that minimize the propagation of compounding pulse errors \cite{ref:Puzzuoli}. 
Note that because ensemble spin inversions commute with the dipolar Hamiltonian, a sequence of AFPs cannot refocus the dephasing caused by homonuclear dipolar interactions.
Therefore, the single-spin transverse magnetization in between AFP applications decays with the corresponding time constant $T_2$.
For a discussion of how the measured signal depends on the delay time ($t_w$) along with supporting experimental data, see Ref. \cite{ref:Supplements}.
We also add an additional free evolution time $(30016-n)t_w$ after each $n$-pulse train, to make sure that $M_z(n)$ for all pulse numbers experience identical $T_1$\nobreakdash-decays.
Although $T_1$ was longer than we could determine with our measurements, we did confirm that it was longer than 5~s.
Fig. \ref{fig:Experimental}(a) shows the resulting normalized $M_z(n)$ signal.
From this data, we wish to extract the ensemble-averaged single-AFP inversion accuracy
\begin{align}
\mathcal{A} = \Big\vert \int_0^\infty d\omega_1~p_1(\omega_1) \zeta_1(\omega_1) \Big\vert,     
\end{align}
where $\zeta_1$ is the single-AFP response function.
We show in Ref. \cite{ref:Supplements} that under certain reasonable assumptions, the signal from our measurement sequence is expected to satisfy $\vert M_z(n) \vert \leq \mathcal{A}^n$.
Consequently, from the exponential fit to $M_z(n)$, we expect that a single AFP inverts spins with an ensemble-averaged accuracy of at least 99.997(3)\% over our Rabi distribution.
To compare the measured fidelity with that of the WURST and Sech/Tanh reference pulses of Sec. \ref{sec:adiabaticMR}, we conducted simulations based on the Lindblad equation that accounts for the $T_2$ dephasing during the wait times (see Ref. \cite{ref:Supplements} for details).
The inhomogeneous broadening that gives rise to the observed $T_2^*$ is accounted for by averaging over a set of Lorentzian-distributed resonance offsets.
We do not model any non-unitary decay process during the AFP itself.
Additionally, our model approximates the effect of all coherent many-body interactions during the wait times with single-spin Lindblad equations.
Hence, the simulations only provide an upper bound on the pulse performance.
A comparison with the calculated ideal fidelity for the optimized reference pulses [inset in Fig. \ref{fig:Experimental}(a)] indicates a significant enhancement in performance for the optimized polynomial AFP.

We investigate the robustness of the pulse to resonance offsets using a train of 5000 AFPs, each separated by $t_w = 52~\mu$s, and measuring $M_z(\delta \omega)$ as we offset the carrier frequency by $\delta \omega/(2\pi)\in[-200~\text{kHz}, 200~\text{kHz}]$ from the center frequency $\omega_0$.
In the detection protocol, we target the same Rabi frequency range [$p_1(\omega_1)$].
The data, presented in Fig. \ref{fig:Experimental}(b), indicates that after 5000 AFPs, the integrated spin signal within the design range $\omega_1 \in [\Omega_1, 2\Omega_1]$ decays to half its peak value in a $\pm 120$ kHz band around the center frequency.
As shown in Fig. \ref{fig:Experimental}(b), our measurements closely track the expectation from simulations.

To characterize the pulse performance as a function of Rabi frequency, we use the same 5000 AFP pulse train, and adjust the Rabi range of the detection protocol [$p_2(\omega_1)$ in Fig. \ref{fig:Experimental}(d)] to be wider than the design range of the AFP.
As before, we add enough free evolution time in the $p_2(\omega_1)$ measurement to compensate for the extra $T_1$ decay in the $\sim 280$ ms-long 5000 AFP measurement sequence.
The resulting $p_2(\omega_1)$ and $p_2(\omega_1) \zeta_m (\omega_1)$ are shown in Fig. \ref{fig:Experimental}(d). We note that $\zeta_m(\omega_1)$ and $\zeta_c(\omega_1)$ correspond to the measured and calculated AFP response functions, respectively.
The calculated response $\zeta_c(\omega_1) = \text{Tr}[\rho(\omega_1)\sigma_z]$ is plotted as the dashed curve in Fig. \ref{fig:Experimental}(d).
The figure clearly indicates that the AFP has the expected performance in the $[\Omega_1, 2 \Omega_1]$ design range, and that the overall Rabi dependence is in excellent agreement with the simulation.
The measured magnetization in this range experiences an average drop by $\sim 15\%$ over the 5000 AFPs, which is consistent with the estimate of $1-e^{-5000/34111}= 13.6\%$ calculated from the ensemble magnetization decay experiment.
Nevertheless, the calculated $\zeta_c(\omega_1)$ shows an average reduction of $\sim 7\%$ over the same $[\Omega_1, 2 \Omega_1]$ Rabi frequency range.
While we do not know the source of the discrepancy, it could arise from small unaccounted perturbations not included in the simulation, such as pulse waveform distortions, coming from the RF electronics, phase noise from the arbitrary waveform generator, or residual spin couplings.
We note that in our case, addressing the transfer function of the electronics in the control searches was not necessary for satisfactory pulse performance.
The Van Loan formalism does, however, allow for the efficient inclusion of transfer function distortions in the optimizations \cite{ref:Haas_2019}.

We conclude the section by comparing our work to some existing experimental results. The performance and robustness of adiabatic control sequences and control sequences derived using shortcuts to adiabaticity techniques have been explored experimentally in the context of nuclear magnetic resonance \cite{ref:Hwang_1998}, Bose-Einstein condensates \cite{ref:Bason_2012} and nitrogen-vacancy centres in diamond \cite{ref:Zhang_2013}. In \cite{ref:Hwang_1998} the authors study the performance of two analytically derived `Tanh/Tan' AFPs, optimized to yield fast broadband inversions, as a function of resonance offset ($\delta \omega$) and Rabi strength ($\omega_1$). The AFPs, when scaled to our target $\omega_1$ values, are $3.7~\mu$s and $5.5~\mu$s-long, with maximum resonance offsets of $7.3$~MHz and $10$~MHz, respectively.
Single-spin simulations, identical in protocol to the ones displayed in Fig.~\ref{fig:Experimental}(a), show that the `Tanh/Tan' AFPs with durations $3.7~\mu$s and $5.5~\mu$s would yield $M_z(n = 5000) = 0.11$ and $M_z(n = 5000) = 0.23$, respectively.
Thus, even though these pulses have maximum resonance offsets that are considerably higher than the 2.3 Rabi cycle pulse [$\Delta\omega_{\text{max}}/(2\pi)=2.5$~MHz], their performance is significantly worse.
Ref. \cite{ref:Bason_2012} characterizes the robustness of a high-fidelity superadiabatic tangent sequence, whose duration, when scaled to our target $\omega_1$ values, is $\sim 7~\mu$s. The authors quote the simulated fidelity of the sequence over the range of $\left[ \Omega_1, 2 \Omega_1 \right]$ as $>0.999$ while the experimental fidelity is confirmed to be $\gtrsim 0.99$ over the same range.

%%% Conclusion %%%
\section{Conclusion \& Outlook}\label{sec:conclusions}
We have developed a numerical control engineering protocol that combines gradient-based optimization with the Van Loan auxiliary matrix formalism \cite{ref:Haas_2019} to provide an efficient and systematic means of designing adiabatic pulses that are robust to a variety of parameter variations and perturbation Hamiltonians.
Using the protocol, we engineered a rapid, 2.3 Rabi cycle-long AFP that addresses the broad Rabi field distribution in our experiments.
The pulse exhibited an infidelity improvement of roughly two orders of magnitude over analytically-derived WURST and Sech/Tanh waveforms of the same duration, by optimizing the Bloch sphere trajectory, and taking advantage of the flexibility of the polynomial ansatz [Eq.(\ref{eq:Ansatz})] for waveform parametrization.

Although the spin evolution clearly diverged from perfect adiabaticity for this short pulse duration, the AFP yielded an experimental inversion accuracy of 99.997\% for the spin ensemble using our broadband RF control electronics.
The engineered AFP exemplifies how the protocol provides a means of addressing the precise demands of a quantum control application, while complying with the constraints and fully utilizing the resources of a particular experimental setup to engineer high-fidelity operations.

With the appendices we provide 3 additional examples, each motivated by particular experiments, further
illustrating the flexibility and utility of the protocol for a range of applications.
We believe that the selective adiabatic control engineering, which we demonstrate with the example in App. \ref{app:slice-selective-AFPS}, is a particularly powerful capability as it provides a way of designing robust state-to-state transfers that are conditional on a specific Hamiltonian parameter.
Such operations provide both a way for selectively addressing specific members of an ensemble that is controlled globally -- a common challenge for sensing and spectroscopy -- as well as a means for characterizing Hamiltonian parameters in the presence of challenging experimental conditions.

Because our protocol builds directly on existing Van~Loan auxiliary matrix methods, it inherits all of the demonstrated capabilities of the former. This includes the ability to engineer robustness with respect to stochastic operators characterized by their power spectral density functions, as well as the use of different basis functions for waveform parametrization, e.g., piece-wise constant functions \cite{ref:Haas_2019}. While our protocol enables the optimization of perturbation expressions using waveform parametrizations that have previously been considered by analytical schemes \cite{ref:Daems_2013}, it is not limited to a specific Hilbert space dimension, nor to particular Hamiltonian generators \cite{ref:Superadiabatic-2016}.
Furthermore, the protocol enables direct implementation of arbitrary control waveform amplitude and bandwidth constraints; such schemes have been developed for engineering unitary pulses using Van~Loan methods in \cite{ref:Haas_2019}, and experimentally demonstrated in \cite{ref:PRX}. The ability to directly include the effect of stochastic operators into adiabatic control optimizations could also prove valuable, as analytical techniques for assessing the effect of particular stochastic operators on various shortcuts to adiabaticity control schemes have been developed in \cite{ref:Ruschhaupt_2012}. Moreover, unlike analytical methods, our protocol enables the use of noise power spectral densities specific to the experimental setup and quantum system at hand.

Because adiabaticity is a frame-dependent measure, the success of our control engineering protocol for the examples shown relied on choosing a favorable reference frame for expressing the spin Hamiltonian in Eq.(\ref{eq:MRHamiltonian}).
The reference frame and the waveform parametrization that we used were inspired by analytical work on adiabatic control \cite{ref:Return}. We found that for the control problems considered here the polynomial ansatz yielded much better convergence to solutions than waveform parametrizations involving smooth time-localized basis functions, such as Gaussians.
We expect that similar reasoning can guide the use of our protocol for quantum systems with different Hilbert space dimensions and system/control Hamiltonians in a variety of quantum control applications requiring high-fidelity state preparations, including quantum computing, simulation, sensing and spectroscopy.

\begin{acknowledgments}
This work was undertaken thanks in part to funding from the U.S. Army Research Office through Grant No.
W911NF1610199, the Canada First Research Excellence
Fund (CFREF), and the Natural Sciences and Engineering Research Council of Canada (NSERC). The University of Waterloo's QNFCF facility was used for this work. This infrastructure would not be possible without the significant contributions of CFREF-TQT, CFI, Industry Canada, the Ontario Ministry of Research and Innovation and Mike and Ophelia Lazaridis. Their support is gratefully acknowledged.
We would like to acknowledge Daniel Puzzuoli for his helpful comments.
\end{acknowledgments}

\appendix

\section{AFP Pulses Robust Against Dipolar Couplings}\label{app:el-AFPS}

We now discuss the design of AFPs for densely-packed electrons that are engineered for robustness against spin-spin dipolar interactions. Such considerations can become relevant for quantum sensing applications; for example with nitrogen vacancy centers, where phase shifts for adiabatic pulses have been used to reduce the effect of nonadiabatic couplings \cite{ref:Genov-2020}.
We optimize two AFPs, one that utilizes an additional perturbation metric to provide built-in robustness against dipolar couplings, and a second AFP that does not include this robustness condition.
The performance of the two pulses are compared in a multi-spin simulation to investigate the effectiveness of the perturbative treatment.

\begin{figure}[ht]
    \includegraphics[width = \columnwidth]{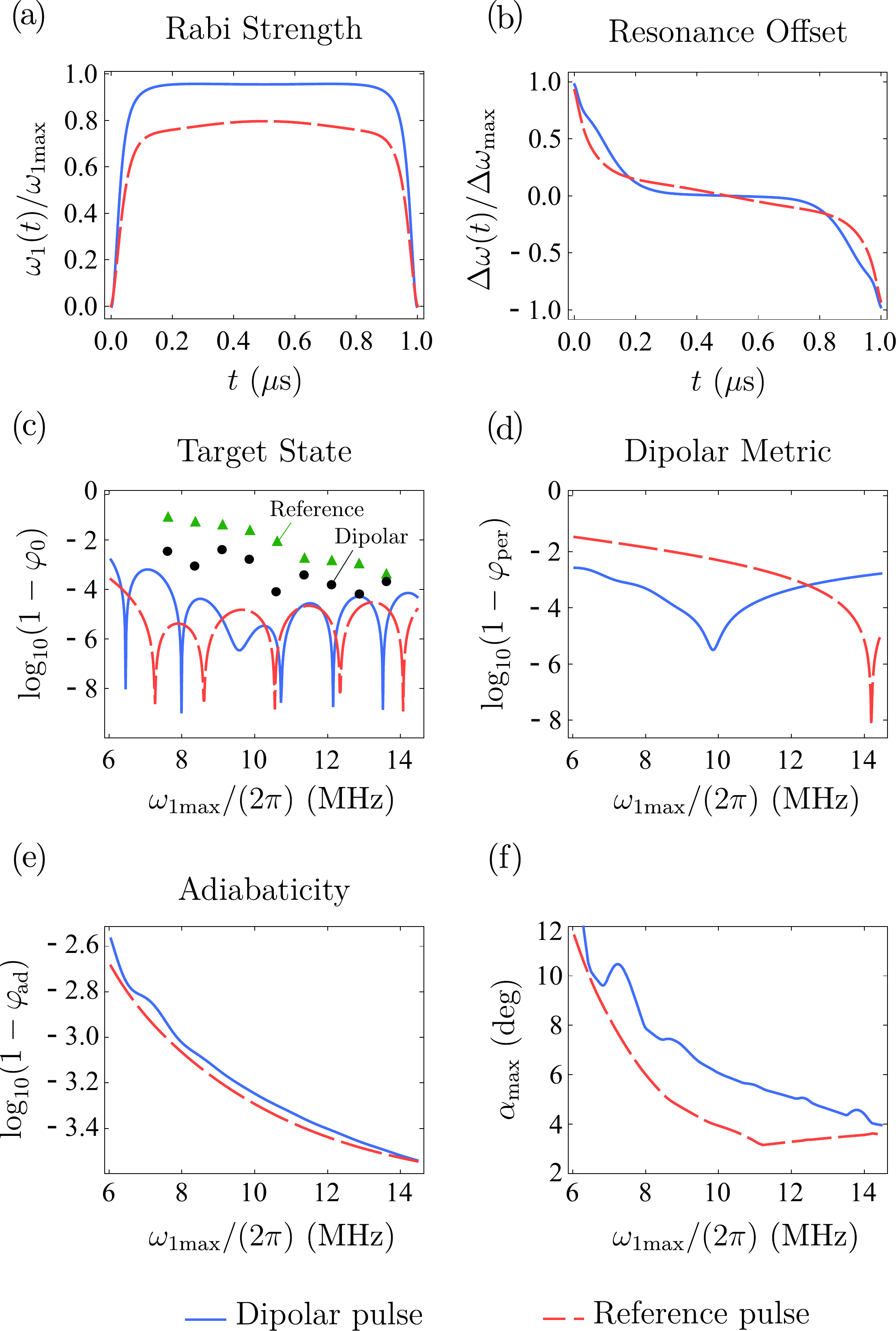}
    \caption{(a) Amplitude and (b) resonance offset waveforms of the dipolar and reference AFPs.
    (c) Target state infidelity of the dipolar and reference pulses (curves) calculated for a single non-interacting spin, along with the mean infidelities from the 7-spin simulation with non-zero dipolar couplings (points).
    (d) Dipolar perturbation metrics for the two pulses. The dipolar metric of the reference pulse is plotted for reference only and was not included in the optimization.
    (e) Adiabaticity metrics for the two pulses.
    (f) The maximum angle between the effective field and magnetization vectors throughout the evolution as a function of Rabi frequency.
        \label{fig:Dipolar} }
\end{figure}

Following the approach of Sec. \ref{sec:adiabaticMR}, the two-body secular dipolar Hamiltonian $\delta H = 2 \sigma_z \otimes \sigma_z-\sigma_x \otimes \sigma_x-\sigma_y \otimes \sigma_y$ \cite{ref:abragam-1983} is used to minimize the leading order correction to the final state $\mathcal{D}_{U\otimes U}( \delta H; T) (\stateI \otimes \stateI)$, which is done by defining a perturbation metric similar to Eq.(\ref{eq:PerturbationMetric}), to be evaluated using a suitable Van Loan generator (see Ref. \cite{ref:Supplements} for details).
The pulse searches are done on an optimization set of 7 electron spins [$\gamma_e/ (2 \pi) = 28.024$ GHz/T], with their maximum Rabi frequency ranging between 7.57 MHz and 12.05 MHz.
The maximum resonance offset of the pulses is set to $\Delta \omega_{\maxx}/(2 \pi) = 50$ MHz, and the duration is chosen to be $T = 1$ $\mu$s.
Both pulses were parametrized using Eq.(\ref{eq:Ansatz}) with $N=40$.
For one of the pulses, which we call the reference pulse, we only use final state and adiabaticity metrics with $p_0^{(\lambda)} = 0.2$ and $p_\ad^{(\lambda)} = 0.8$.
For the dipolar pulse, the metric coefficients are chosen as $p_0^{(\lambda)} = 0.2$, $p_\ad^{(\lambda)} = 0.5$ and $p_\pert^{(\lambda)} = 0.3$.
The weights between different set members are $w^{(\lambda)} =1/7$ for both pulses, and the rest of the optimization is performed in the same way as before. The resulting pulses and their associated control metrics are shown in Fig. \ref{fig:Dipolar}.

To examine the effect of dipolar couplings, we simulate a system of 7 interacting electrons experiencing a particular Rabi frequency for 9 different Rabi frequency values.
The spatial coordinates of the 7 electrons were chosen by taking a 4 nm-sided cube, arranging spins on its center and the centers of its faces, and then displacing each spin by a random vector drawn from a uniform distribution on $[-0.5\text{ nm},0.5\text{ nm}]^3$.
For reference, the largest dipolar coefficient for the spatial arrangement is $6.5$ MHz.
Starting from the state $\spinup$ for each spin, we compute the mean fidelity $\varphi = \sum_{j=1}^7 \text{Tr}[\rho_j(T)\spindown \spindownd]/7$ for the 9 different Rabi frequencies, where $\rho_j(T)$ is the final reduced density matrix of the $j$th spin.
The results for both the dipolar and reference pulses, along with the fidelity metric $\varphi_0$, which is calculated for a single non-interacting spin, are shown in Fig. \ref{fig:Dipolar}(c).
The plots show that despite the reference AFP exhibiting slightly better target state and adiabaticity metrics for a single spin, the mean infidelity of the dipolar AFP computed for the coupled 7-spin system is approximately an order of magnitude smaller, in the AFP design range.

Finally, the performance of the reference pulse improves as the Rabi frequency is increased to $\sim 14$ MHz, which is consistent with the Rabi frequency dependence of its dipolar metric [Fig. \ref{fig:Dipolar}(d)].
This implies that the main contributor to its inferior performance is the absence of the dipolar metric in the pulse search.
We therefore expect the ability to minimize arbitrary perturbation terms to be a powerful tool for engineering fast adiabatic pulses robust to interactions.

\section{Larmor-Selective AFP Pulses} \label{app:slice-selective-AFPS}
Quantum control operations conditional on some Hamiltonian parameter for a globally controlled ensemble of quantum systems underpin all spectroscopy and most imaging applications \cite{ref:Degen}, and are often a key ingredient for quantum sensing \cite{Degen-2017, Barry-2020}. Here, we demonstrate the use of our control protocol for engineering adiabatic operations that are conditional on a single Hamiltonian parameter. Motivated by the nano-MRI experiments in \cite{ref:Degen}, we demonstrate optimizing controls that adiabatically invert spins experiencing a certain adjustable range of Larmor frequencies, called the inversion band, while not inverting spins outside of that range. We use the same experimental parameters as in \cite{ref:Degen}.

Using the polynomial ansatz of Eq.(\ref{eq:Ansatz}) with $N = 10$, we design a $300 \ \mu$s-long AFP for proton spins with gyromagnetic ratio $\gamma/(2 \pi) = 42.577$ MHz/T, under a maximum Rabi frequency of $\omega_{1\maxx}/(2 \pi) = 225.7$~kHz, and a maximum resonance offset of $\Delta \omega_\maxx/(2 \pi) = 75$ kHz.
We search for Larmor-selective pulses by considering an optimization set of spins $(\Gamma)$ at different resonance offsets, and assigning different single-member target functions depending on whether the the optimization set member ($\lambda \in \Gamma$) is inside or outside the inversion band.
Taking the initial state to be $\spinup$, for $\lambda$ inside the inversion band, we assign the metric coefficients $p_0^{(\lambda)} = 0.2$ and $p_\ad^{(\lambda)} = 0.8$, with the target state being set to $\spindown$; whereas the $\lambda$ outside the inversion band have $p_0^{(\lambda)} = 1$ and $p_\ad^{(\lambda)} = 0$, with a target state equal to the initial state.
Within the inversion band range of $[-47 \text{ kHz}, 47\text{ kHz}]$, we use 11 distinct $\lambda \in \Gamma$, with additional 2 members of the optimization set outside the band at $\pm 72.5$ kHz. We reinforce the sharpness of the band edges in the optimization by choosing $w^{(\lambda)} = 2/17$ for the $\pm 47$ kHz and $\pm 72.5$ kHz members, and $w^{(\lambda)}=1/17$ for the others.

\begin{figure}[ht]
    \includegraphics[width = \columnwidth]{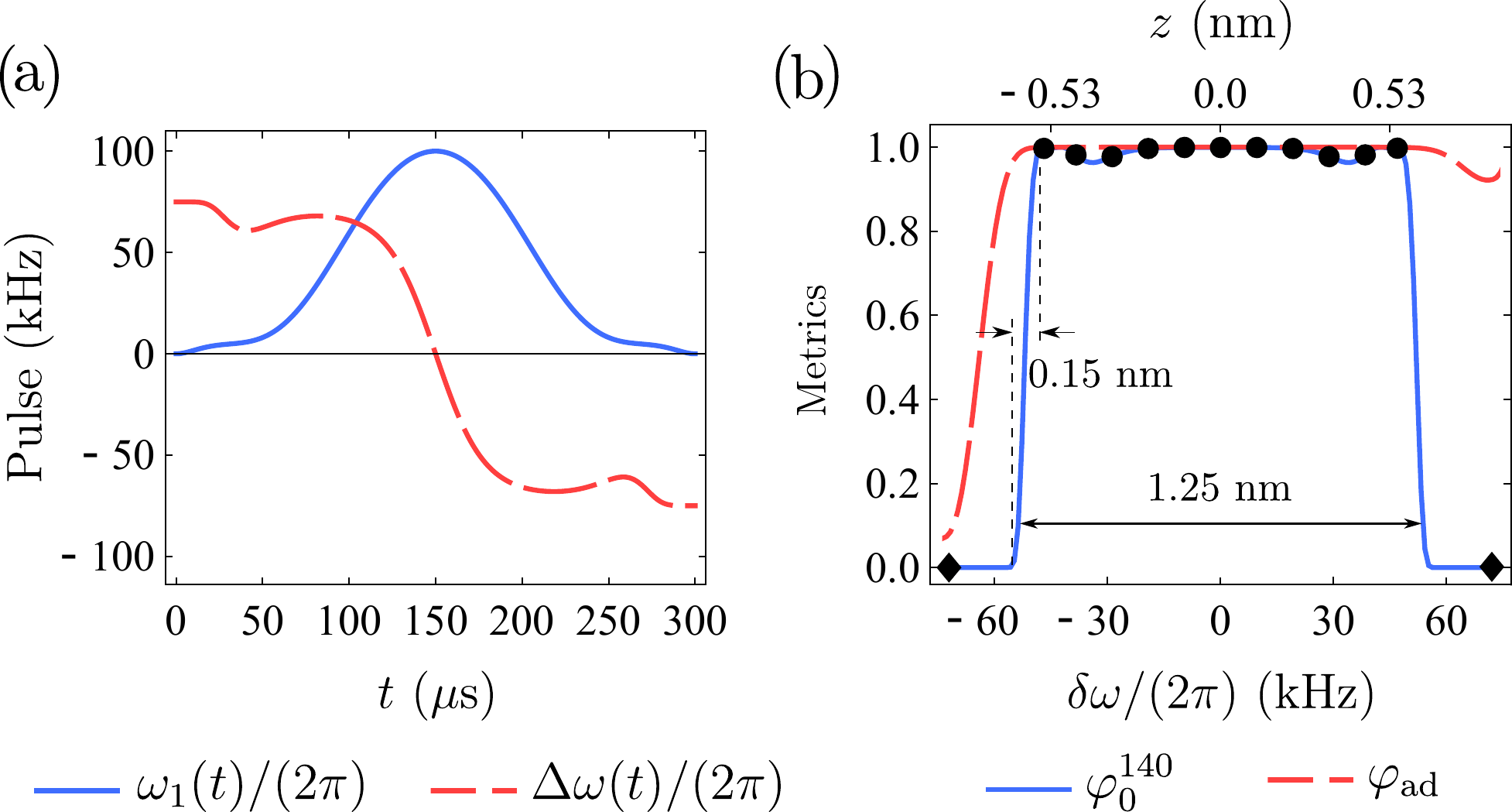}
    \caption{(a) Amplitude and resonance offset waveforms for the optimized Larmor-selective AFP.
    (b) Adiabaticity metric $\varphi_\ad$ and target state fidelity of the pulse $\varphi_0^M$ after $M=140$ spin inversions, as a function of detuning.
    The $z = \delta \omega/(\gamma G)$ axis was determined using a static field gradient of $G=2\times 10^6$ T/m.
    The arrows indicate the width and edges of the inversion band. The markers in subfigure (b) represent the members of the optimization set $\lambda \in \Gamma$, with the circles (diamonds) being inside (outside) the inversion band.
        \label{fig:SliceSelection}
    }
\end{figure}

The optimized pulse shape is given in Fig. \ref{fig:SliceSelection}(a). Just like \cite{ref:Degen}, we look at the fidelity metric $(\varphi_0)^M$ as a function of detuning $\delta \omega/(2 \pi)$, where $M=140$ is the number of adiabatic inversions applied in the experiment. Fig. \ref{fig:SliceSelection}(b) shows the profile of $\varphi_0^M$, for which the total band width, defined by the region in which $\varphi_0^M \geq 0.1$, equals 107.8 kHz. 
For a static field gradient of $G = 2\times 10^6$~T/m utilized in \cite{ref:Degen}, this translates to a near-nanometer spatial band width of $\delta z = \delta \omega/(\gamma G) =1.25$ nm.
The band edge width is determined to be $\sim 13$ kHz or 0.15 nm.
The pulse also generates the expected adiabatic behavior, as the maximum angle between the effective field and magnetization vectors in the $[-47 \text{ kHz}, 47\text{ kHz}]$ range is $4^\circ$, while the angle at the edges ($\pm$ 47 kHz) reaches $\sim 15^\circ$.

We have thus demonstrated an example for systematically engineering adiabatic operations conditional on a Hamiltonian parameter suitable for nanometer-scale MRI experiments. We note that if required, robustness to Rabi field variations can also be directly addressed in the optimization by considering a perturbation Hamiltonian {$\delta H(\xx,t) = b_x(\xx,t) \sigma_x$}, and minimizing the associated metric.

\section{Adiabatic Operation Connecting Arbitrary States}\label{app:arbitraryState}
Eq.(\ref{eq:Ansatz}) restricts the initial and final effective fields to the $z$ axis. For arbitrary adiabatic spin 1/2 state-to-state transfers -- a useful capability for experiments
\cite{ref:Yang_2019, Saywell-2020} -- a more general parametrization is needed \cite{ref:Return,ref:Yang_2019, ref:Zlatanov-2020}.
To this end, we first look for a polynomial function that connects the arbitrary points $(0,\xi)$ and $(T,\xi')$.
One solution is to use the line connecting these two points, and add it to the most general polynomial with roots at $t \in \{0,T\}$.
We thus define the function
\begin{eqnarray}
    f^{\xi \xi'}_{mm'}(\xx,t) \equiv \frac{t}{T} (1- \frac{t}{T}) && \sum_{n=m+1}^{m'} x_n \big(1-2 \frac{t}{T}\big)^{n-m-1}  \nonumber\\  &&  \qquad \quad + \frac{t}{T} (\xi' - \xi) + \xi, \label{eq:polFunc}
\end{eqnarray}
where $\xi = f^{\xi \xi'}_{mm'}(\xx,0)$ and $\xi' = f^{\xi \xi'}_{mm'}(\xx,T)$.
One can now use polynomials of this form to parametrize the three effective field components.
To also incorporate amplitude and bandwidth limitations, we utilize tangent hyperbolic soft clipping functions, which leads to the ansatz
\begin{equation}
    \begin{cases}
    \displaystyle b_x(\xx,t) = \frac{\omega_{1\maxx}}{\sqrt{2}} \tanh\big[f_{0,N/3}^{\alpha \alpha'} (\xx,t)\big]\\[10pt]
    \displaystyle b_y(\xx,t) = \frac{\omega_{1\maxx}}{\sqrt{2}} \tanh\big[f_{N/3,2N/3}^{\beta \beta'} (\xx,t)\big]\\[10pt]
    \displaystyle b_z(\xx,t) =\Delta \omega_{\maxx} \tanh\big[f_{2N/3,N}^{\gamma \gamma'} (\xx,t)\big]
    \end{cases}, \label{eq:AnsatzGeneral}
\end{equation}
where $N$ is the number of optimization parameters. The $(\alpha,\beta,\gamma)$ and $(\alpha',\beta',\gamma')$ indices in Eq.(\ref{eq:AnsatzGeneral}) are chosen as
\begin{center}
\begin{tabular}{ l l }
 $\alpha=\tanh^{-1} n_x,$  & $\alpha'=\tanh^{-1} n_x',$ \\[4pt]
 $\beta=\tanh^{-1} n_y,$  & $\beta'=\tanh^{-1} n_y',$  \\[4pt]
 $\displaystyle\gamma= \tanh^{-1} \big(\frac{\omega_{1\maxx}}{\sqrt{2} \Delta \omega_\maxx}n_z\big),$ & $\displaystyle\gamma'= \tanh^{-1} \big(\frac{\omega_{1\maxx}}{\sqrt{2} \Delta \omega_\maxx}n_z'\big),$ 
\end{tabular}
\end{center}
such that for the design Rabi frequency, $\bb(\xx,0)\propto \hat{\mathbf{n}}$ and $\bb(\xx,T)\propto \hat{\mathbf{n}}'$, where $\hat{\mathbf{n}} = (n_x,n_y,n_z)$ and $\hat{\mathbf{n}}' = (n_x',n_y',n_z')$ are unit vectors along the initial and final states on the Bloch sphere, respectively.

\begin{figure}[ht]
    \centering
    \includegraphics[width =\columnwidth]{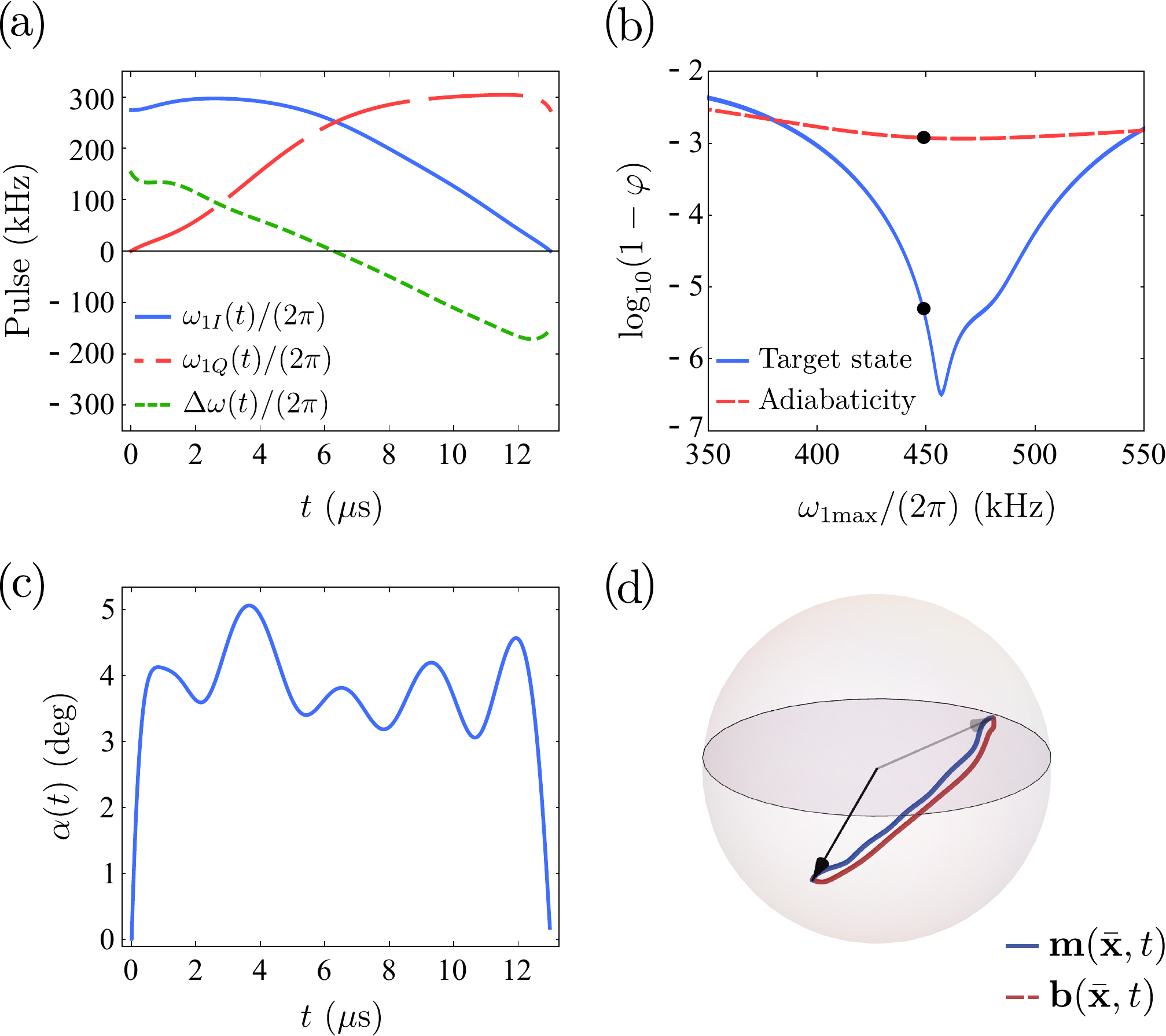}
    \caption{(a) Adiabatic pulse designed for evolving the $(\vartheta_i,\psi_i) = (\pi/3,0)$ point to $(\vartheta_f,\psi_f) = (2\pi/3,\pi/2)$ on the Bloch sphere. The field values correspond to the spin in the optimization set.
    (b) The target state and adiabaticity metric infidelities as a function of Rabi strength. The dot indicates the spin used in the optimization. 
    (c) The maximum angle between the effective field and magnetization vectors for the optimized $\omega_{1\maxx}/(2 \pi) = 448$~kHz spin as a function of time, and (d) the corresponding Bloch sphere trajectory of the magnetization and effective field.}
    \label{fig:ArbitraryState}
\end{figure}

We now give a brief example of an operation that transfers an arbitrary state on the Bloch sphere to another arbitrary state. 
For this we design a $T=13 \ \mu\text{s}$ pulse for spins with $N=30$ control parameters.
We enforce the bandwidth and amplitude constraints $\Delta \omega_\maxx/(2\pi) = 7.4$ MHz and $\omega_{1 \maxx}/(2 \pi) = 448$~kHz on a single-member optimization set, with target function coefficients $p_0 = 0.2$ and $p_\ad = 0.8$, and no perturbation metric in mind.
The initial and final states for the pulse design are set to $(\vartheta_i,\psi_i) = (\pi/3,0)$ and $(\vartheta_f,\psi_f) = (2\pi/3,\pi/2)$, respectively, where $\vartheta$ and $\psi$ denote the polar and azimuthal angles of the states on the Bloch sphere, respectively.
The resulting pulse, along with its associated performance metrics and calculated Bloch sphere trajectory are presented in Fig. \ref{fig:ArbitraryState}.
The results show that the spin follows the effective field with an angle $\alpha(t) \leq 5^\circ$, and reaches the target state with a fidelity higher than 0.99999.
The target state infidelity [Fig. \ref{fig:ArbitraryState}(b)] exhibits a single minimum near $\omega_{1 \maxx}/(2 \pi) = 448$~kHz, as deviations from the design Rabi frequency will necessarily cause misalignment between the initial state and effective field.
The small shift between the minimum and the design Rabi frequency is due to coherent effects arising from the slight non-adiabaticity of the evolution. 

\nocite{*}
%\bibliography{refs}{}% Produces the bibliography via BibTeX.

\providecommand{\noopsort}[1]{}\providecommand{\singleletter}[1]{#1}%
%

%%% Supplements %%%
\clearpage


\section*{Supplementary material}

# Supplemental Material for Numerical Engineering of Robust Adiabatic Operations

Sahand Tabatabaei,[1,2] Holger Haas,[1,2] William Rose,[3] Ben Yager,[1,2] Michèle Piscitelli,[1,2]
Pardis Sahafi,[1,2] Andrew Jordan,[1,2] Philip J. Poole,[4] Dan Dalacu,[4] and Raffi Budakian[1,2,5,∗]

[1]*Department of Physics, University of Waterloo, Waterloo, ON, Canada, N2L3G1*

[2]*Institute for Quantum Computing, University of Waterloo, Waterloo, ON, Canada, N2L3G1*

[3]*Department of Physics, University of Illinois at Urbana-Champaign, Urbana, Illinois 61801, USA*

[4]*National Research Council of Canada, Ottawa, Ontario, Canada, K1A 0R6*

[5]*Canadian Institute for Advanced Research, Toronto, ON, Canada, M5G1Z8*

## I. EXTENSION TO MULTI-LEVEL SYSTEMS: AN EXAMPLE

Our definitions for the adiabaticity metric $\varphi_{\mathrm{ad}}(\mathbf{x}) = \langle\psi_0|U^\dagger(\mathbf{x}, T)\mathcal{D}_U(iP; T)|\psi_0\rangle/T$ and the final state metric $\varphi_0(\mathbf{x}) = |\langle\psi_T|U(\mathbf{x}, T)|\psi_0\rangle|^2$ in the main text are equally applicable to multi-level systems, provided that their eigen-projection operator $P$, along with its derivatives (for the $\boldsymbol{\nabla}\varphi$ computation), are known as a function of time and the controls. The Van Loan generator needed for the metric calculation is exactly the same as Eq.(7) of the main text.

An example of such a problem is the adiabatic control of a general spin-$s$ nucleus in magnetic resonance. In this case, the Hamiltonian has the form [1]

$$H(\mathbf{b}(\mathbf{x}, t)) = -\mathbf{b}(\mathbf{x}, t) \cdot \mathbf{S}, \tag{S1}$$

with $\mathbf{b}(\mathbf{x}, t)$ being the effective field, and $\mathbf{S}$ being the spin vector operator. The Hamiltonian enables certain, albeit not all, state-to-state transfers. We note that Eq.(S1) is written with the assumption that the spin does not experience quadrupolar coupling due to an electric field gradient [2], which is possible in cases where it is suppressed due to certain symmetries of the sample. This treatment is also applicable to systems with a small quadrupolar coupling relative to the effective field, such that it can be incorporated using an additional perturbation metric. Denoting the eigenbasis of $S_z$ by $\{|m\rangle\}_{m=-s}^s$, The eigenstates of $H(\mathbf{b})$ can be derived by rotating the coordinate frame into one in which the effective field coincides with the $z$ axis. Expressing the effective field $\mathbf{b}(\mathbf{x}, t)$ in spherical coordinates, this coordinate transformation corresponds to a rotation around the $y$ axis by the polar angle $\vartheta$, followed by a rotation around the (old) $z$ axis by the azimuthal angle $\psi$. Writing the rotated states in terms of the Wigner small d-matrix elements [3] $d_{m'm}^{(s)}(\vartheta) \equiv \langle m'|e^{-i\vartheta S_y}|m\rangle$ results in the eigenstates

$$|E_m(\mathbf{b})\rangle = \sum_{m'=-s}^s e^{-im'\psi}d_{m'm}^{(s)}(\vartheta)|m'\rangle. \tag{S2}$$

The Wigner d-matrix elements in Eq.(S2) can be looked up from a standard reference (e.g. [3]) for a particular spin number $s$, and used to construct the eigenprojection operators

$$
\begin{aligned}
P_m(\mathbf{b}) &= |E_m(\mathbf{b})\rangle\langle E_m(\mathbf{b})| \\
&= \sum_{m_1=-s}^s \sum_{m_2=-s}^s e^{-i(m_1-m_2)\psi} \, d_{m_1,m}^{(s)}(\vartheta)\left(d_{m_2,m}^{(s)}(\vartheta)\right)^* |m_1\rangle\langle m_2|.
\end{aligned}
\tag{S3}
$$

One can thus use the exact same procedure described for two-level systems, and an appropriate parametrization of the effective field, to engineer permissible adiabatic operations for spin $s > 1/2$ nuclei in MR.

---

∗ rbudakian@uwaterloo.ca

## II.  POLYNOMIAL COEFFICIENTS FOR THE 2.3 RABI-CYCLE AFP

The optimized coefficients $\bar{\mathbf{x}} \in \mathbb{R}^{50}$ of the 2.3 Rabi-cycle polynomial AFP discussed in Sections IV-V of the main text are given as

$$
\begin{aligned}
\bar{\mathbf{x}} = (&7.023645,\ 6.839118,\ 3.854925,\ 4.158147,\ 2.552735,\ 1.309126,\ -0.668767,\ -2.081916,\ -0.710703,\ -1.766848, \\
&-1.709370,\ -1.923206,\ -1.610771,\ -2.548112,\ -2.020481,\ -2.236042,\ -0.554534,\ -0.287112,\ 0.388593, \\
&0.318120,\ 1.518851,\ 1.420478,\ 0.645996,\ 1.424078,\ 2.268813,\ 0.270238,\ 0.202866,\ -0.095743,\ 0.992899, \\
&-0.278452,\ -0.209323,\ -0.64334,\ -0.300252,\ 1.231855,\ 1.271526,\ 0.967405,\ 0.999899,\ 0.785292,\ 1.551433, \\
&1.112399,\ 1.554735,\ 1.626168,\ -0.237164,\ 0.868774,\ 0.625801,\ 1.1631,\ 0.587447,\ 0.603619,\ 0.16885, \\
&-0.719233)
\end{aligned}
$$

$$(\text{S4})$$

The analytic form of the pulse shape can thus be determined using these values and the polynomial ansatz [Eq.(12)] given in the main text.

## III.  REFERENCE PULSE OPTIMIZATION

In this section we discuss the optimization details for the WURST and Sech/Tanh pulses used in Section IV of the main text as reference pulses for benchmarking the 2.3 Rabi cycle polynomial AFP.

The WURST pulse shape [4] is parametrized with three variables, the maximum amplitude $A \in [0,1]$, modulation depth $\delta \in [0,1]$ and the order $n \in \mathbb{N}$. The associated effective field trajectory for a pulse of duration $T$ is

$$
\mathbf{b}(A, \delta, n, t) = \left( \omega_{1\max} A \left[ 1 - \left| \cos\left( \frac{\pi t}{T} \right) \right|^n \right],\ 0,\ \Delta\omega_{\max} \delta \ \left( 1 - \frac{2t}{T} \right) \right),
$$

$$(\text{S5})$$

i.e. the pulse has a raised cosine amplitude profile with a linear frequency sweep. Similarly, for the Sech/Tanh pulse shape [5] we have

$$
\mathbf{b}(A, \delta, \kappa, t) = \left( \omega_{1\max} A\ \text{sech}\left[ (1 - \frac{2t}{T})\ \text{sech}^{-1}\kappa \right],\ 0,\ \Delta\omega_{\max}\delta \tanh\left[ (1 - \frac{2t}{T})\ \text{sech}^{-1}\kappa \right] \right),
$$

$$(\text{S6})$$

which is also parametrized by the same variables $A$ and $\delta$, along with a truncation factor $\kappa$ that indicates the fractional amplitude of the Rabi field at the start and end of the pulse.

To optimize both reference pulses, we use the exact same Van Loan generator, pulse length, target function coefficients and optimization set used in Section IV of the main text for optimizing the 2.3 Rabi cycle AFP. Since each waveform contains only 3 control parameters, we use the gradient-free constrained optimization algorithm available in Mathematica's NMaximize function [6]. Both pulses are optimized with the same Rabi strength $\omega_{1\max}$ and resonance offset $\Delta\omega_{\max}$ constraints as the polynomial AFP. The order of the WURST pulse was optimized over the set of natural numbers, while the other parameters were constrained to $0 \leq A \leq 1$, $0 \leq \delta \leq 1$ and $0 \leq \kappa \leq 0.1$ in the algorithm. The resulting optimum pulse parameters are provided in Table I.

TABLE I. Optimum reference pulse parameters for benchmarking.

| WURST | $A = 1$ | $\delta = 0.524$ | $n = 3$ |
|---|---|---|---|
| Sech/Tanh | $A = 1$ | $\delta = 0.183$ | $\kappa = 0.077$ |

## IV.  LINDBLAD SIMULATIONS

For the simulations in Section V of the main text, we consider a single spin evolving under an FM-frame effective field $\mathbf{b}(t)$, while experiencing transverse dephasing with time constant $T_d$. Here, the FM-frame Lindblad equation is deduced from the corresponding lab frame equation. The Lindblad equation describing the evolution in the lab frame is [7]

$$
\frac{d}{dt}\tilde{\rho}(t) = -i\left[ \tilde{H}(t), \tilde{\rho}(t) \right] + \frac{1}{2T_d}\left( \sigma_z \tilde{\rho}(t) \sigma_z - \tilde{\rho}(t) \right),
$$

$$(\text{S7})$$

where $\tilde{\rho}(t)$ is the lab-frame density matrix and

$$\tilde{H}(t) = -\omega_0 \frac{\sigma_z}{2} - 2b_x(t) \cos\left((\omega_0 - \Delta\omega)t - \int_0^t dt'\, b_z(t')\right) \frac{\sigma_x}{2},$$

(S8)

is the lab-frame Hamiltonian, with $\Delta\omega$ being an additional constant resonance offset. We transform to the FM frame using the unitary $V(t) \equiv \exp\left[i \int_0^t dt'(\omega_0 - \Delta\omega - b_z(t'))\sigma_z/2\right]$. From Eq.(S7,S8), we derive the following Lindblad equation for the FM frame density matrix $\rho(t) = V^\dagger(t)\tilde{\rho}(t)V(t)$:

$$\frac{d}{dt}\rho(t) = \mathcal{L}_t(\rho(t)),$$

(S9)

where, under the rotating wave approximation, the superoperator $\mathcal{L}_t$ is

$$\mathcal{L}_t(\rho) \equiv i\left[\mathbf{b}(t) \cdot \frac{\boldsymbol{\sigma}}{2} + \Delta\omega \frac{\sigma_z}{2},\ \rho\right] + \frac{1}{2T_d}(\sigma_z \rho \sigma_z - \rho).$$

(S10)

The solution to Eq.(S9) is given by $\rho(t) = \mathcal{E}_t(\rho(0))$, with $\mathcal{E}_t = \text{Texp}\left[\int_0^t dt' \mathcal{L}_{t'}\right]$ given a specific matrix representation of $\mathcal{L}_{t'}$. For this work, we calculate the dynamical map $\mathcal{E}_t$ using its matrix representation in the $\{\frac{1}{\sqrt{2}}, \frac{\sigma_x}{\sqrt{2}}, \frac{\sigma_y}{\sqrt{2}}, \frac{\sigma_z}{\sqrt{2}}\}$, i.e. Bloch basis

$$\text{Texp}\left(\int_0^t dt' \begin{bmatrix} 0 & 0 & 0 & 0 \\ 0 & -1/T_d & b_z(t') + \Delta\omega & 0 \\ 0 & -b_z(t') - \Delta\omega & -1/T_d & b_x(t') \\ 0 & 0 & -b_x(t') & 0 \end{bmatrix}\right),$$

(S11)

which is computed by solving the associated initial value problem using Mathematica's NDSolve function. Note that due to the block-diagonal structure of Eq.(S11), we only need to consider the inner $3 \times 3$ block for computations.

To simulate our three measurement scenarios, we consider a primitive sequence comprised of the 2.3 Rabi-cycle polynomial AFP applied with maximum Rabi frequency $\omega_1$, followed by a wait time $t_w$ in which the spin experiences dephasing with $T_d = T_2 = 364\ \mu s$ and no control sequence ($b_x = b_z = 0$). Our memoryless Lindblad modeling of the dephasing process is justified because spin ensemble inversion operations do not refocus dipolar evolution, which is the main source of dephasing in our sample. We assume no dephasing ($T_d \to \infty$) during the AFP evolution time. The dynamical map under this primitive sequence is $\mathcal{E}(\omega_1, \Delta\omega) = \mathcal{E}_w(\Delta\omega, T_d = T_2)\mathcal{E}_{\text{AFP}}(\omega_1, \Delta\omega, T_d \to \infty)$, with $\mathcal{E}_{\text{AFP}}(\omega_1, \Delta\omega, T_d \to \infty)$ and $\mathcal{E}_w(\Delta\omega, T_d = T_2)$ being dynamical maps for the AFP and wait time evolution, respectively. The 5000-AFP response function $\zeta_c(\omega_1)$ can thus be calculated as

$$\zeta_c(\omega_1) = \sum_{\delta\omega_{\text{inh}}} p_{\text{inh}}(\delta\omega_{\text{inh}}) \text{Tr}\left[\sigma_z \mathcal{E}^{5000}(\omega_1, \delta\omega_{\text{inh}})(|\uparrow\rangle\langle\uparrow|)\right],$$

(S12)

where we have accounted for the $T_2^* = 70\ \mu s$ inhomogeneous broadening by averaging over the Lorentzian line $p_{\text{inh}}(\delta\omega_{\text{inh}}) \propto [1 + (T_2^* \delta\omega_{\text{inh}})^2]^{-1}$, which is normalized such that $\sum_{\delta\omega_{\text{inh}}} p_{\text{inh}}(\delta\omega_{\text{inh}}) = 1$. The sum over $\delta\omega_{\text{inh}}$ in the above equation samples the range $[-20\ \text{kHz}, 20\ \text{kHz}]$ at 21 points.

We simulate the fidelity measurement for $n$ AFPs by calculating

$$M_z(n) = \sum_{\omega_1} \sum_{\delta\omega_{\text{inh}}} p_{1\text{norm}}(\omega_1) p_{\text{inh}}(\delta\omega_{\text{inh}}) \text{Tr}\left[\sigma_z \mathcal{E}^n(\omega_1, \delta\omega_{\text{inh}})(|\uparrow\rangle\langle\uparrow|)\right],$$

(S13)

where $p_{1\text{norm}}(\omega_1) = p_1(\omega_1)/\sum_{\omega_1} p_1(\omega_1)$ is the normalized $p_1$ distribution measured in Section V of the main text, with 27 Rabi-frequency samples taken over $\omega_1 \in [450\ \text{kHz}, 875\ \text{kHz}]$. The $p_{1\text{norm}}(\omega_1)$ distribution is assumed to be constant inside every measured Rabi-frequency bin. We compute the resonance offset decay curve for 5000 AFPs in a similar fashion as a function of the carrier frequency offset $\delta\omega$:

$$M_z(\delta\omega) = \sum_{\omega_1} \sum_{\delta\omega_{\text{inh}}} p_{1\text{norm}}(\omega_1) p_{\text{inh}}(\delta\omega_{\text{inh}}) \text{Tr}\left[\sigma_z \mathcal{E}^{5000}(\omega_1, \delta\omega + \delta\omega_{\text{inh}})(|\uparrow\rangle\langle\uparrow|)\right].$$

(S14)

## V. SIGNAL DEPENDENCE ON THE DEPHASING TIME

We numerically investigate the dependence of the spin signal on the wait time $t_w$ between the AFPs and the wait-time dephasing time constant $T_d$ using the Lindblad simulations discussed in Section IV. It is easy to see by direct calculation of Eq.(S11) that the wait-time dynamical map for zero resonance offset $\mathcal{E}_w(0)$ only depends on the ratio $t_w/T_d$. We verified that including the inhomogeneous line shape only slightly changed the $M_z(n)$ curves, while significantly increasing the simulation time, and did not change the qualitative dependence of spin signal on $t_w$ and $T_d$. Therefore, we do not average over the inhomogeneous line shape in this section, and calculate $M_z(n)$ for $n \in [1, 50000]$ AFP applications for various $t_w/T_d$ values. The result is depicted in Fig. S1, indicating that the longitudinal $z$-magnetization monotonically increases with longer wait times, or equivalently, shorter dephasing time constants. We further verify this behavior experimentally by measuring the spin signal for a sequence of 5000 AFPs separated by wait times $t_w \in \{5~\mu s, 52~\mu s, 200~\mu s, 600~\mu s\}$. The result (Fig. S2) shows very good agreement with the expected trend. We attribute the additional decay relative to simulation for long wait times to unaccounted perturbations such as transfer function distortions and any non-unitary processes during the pulses. Because of this, the measured signal at $t_w = 52~\mu s$ deviates negligibly from points at long wait times.

The signal-dephasing time dependence can be understood through the following simple argument: For each Rabi frequency $\omega_1$, if we assume that the predominant error in implementing a single AFP is either an under or over-rotation by an infinitesimal yet finite angle $\delta\theta(\omega_1)$, then we would evaluate a single AFP fidelity as $\sim \cos\left(\delta\theta(\omega_1)\right) = 1 - \frac{(\delta\theta(\omega_1))^2}{2} + \dots$. If we apply $n$ of these pulses with no wait times in between ($t_w = 0$) the resulting final longitudinal magnetization would evaluate to $\sim \cos\left(n\delta\theta(\omega_1)\right) = 1 - \frac{(n\delta\theta(\omega_1))^2}{2} + \dots$, whereas in the case of full dephasing ($t_w \to \infty$) in between each of the pulses applied, the $z$-magnetization becomes $\sim \cos^n\left(\delta\theta(\omega_1)\right) = 1 - \frac{n(\delta\theta(\omega_1))^2}{2} + \dots$. Since $\cos^n\left(\delta\theta(\omega_1)\right) - \cos\left(n\delta\theta(\omega_1)\right) = \frac{n(n-1)}{2}(\delta\theta(\omega_1))^2 + \dots$, it demonstrates that allowing the transverse magnetization to dephase between AFPs will result in a larger spin signal, especially for a high number of pulse applications. Moreover, using $t_w \to \infty$ also ensures that for a single Rabi frequency, the decay in $z$-magnetization as a function of $n$ approximates an exponential decay in the single-AFP response function $\zeta_1$, i.e. the $n$-AFP response function is $\cos^n\left(\delta\theta(\omega_1)\right) = [\zeta_1(\omega_1)]^n$.

For the experimental verification of the 2.3 Rabi-cycle AFP (Section V of the main text), we chose $t_w = 52~\mu s$, which is the longest possible wait time within the restrictions of our arbitrary waveform generator for the maximum number of applied pulses. This corresponds to a normalized dephasing time $t_w/T_2 = 0.147$, for which the simulated decay curve is also plotted in Fig. S1.

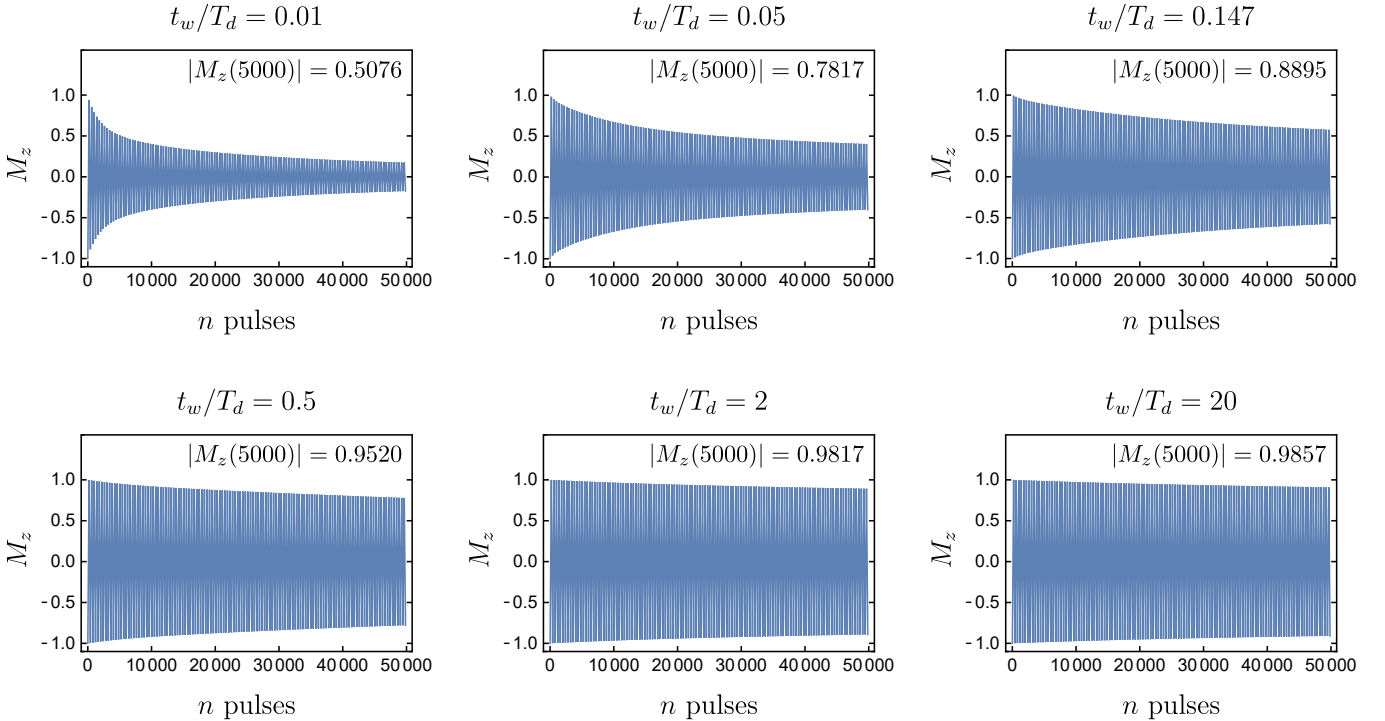

FIG. S1. Simulated spin signal as a function of pulse number $n$, for various normalized dephasing times $t_w/T_d$.

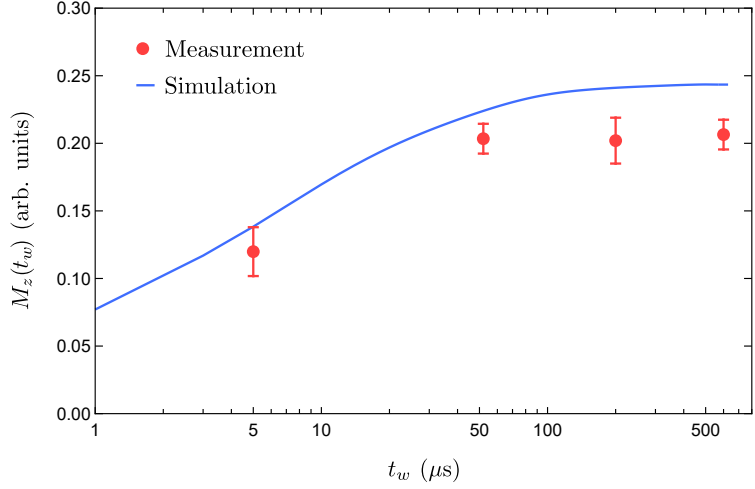

FIG. S2. Measured and simulated spin signal as a function of the wait time $t_w$ for 5000 AFPs.

## VI. LOWER BOUND ON THE SINGLE-AFP INVERSION ACCURACY

We show that the measured signal for $n$ pulses $M_z(n)$ in our fidelity measurement scheme satisfies $|M_z(n)| \leq \mathcal{A}^n$, where

$$\mathcal{A} = \Big| \int_0^\infty d\omega_1 \; p_1(\omega_1)\zeta_1(\omega_1) \Big| = -\int_0^\infty d\omega_1 \; p_1(\omega_1)\zeta_1(\omega_1), \tag{S15}$$

is the ensemble-averaged single-AFP inversion accuracy and $\zeta_1$ is the single-AFP response function. First consider the case of full dephasing $t_w \to \infty$. As shown in Section V, the $n$-AFP response function is $[\zeta_1(\omega_1)]^n$. Now since $\zeta_1(\omega_1) \leq 0$ over the support of $p_1$, we have

$$\begin{aligned}
|M_z(n, t_w \to \infty)| &= \int_0^\infty d\omega_1 \; p_1(\omega_1)\big[-\zeta_1(\omega_1)\big]^n \\
&\leq \Big( \int_0^\infty d\omega_1 \; p_1(\omega_1)\big[-\zeta_1(\omega_1)\big] \Big)^n,
\end{aligned} \tag{S16}$$

meaning that

$$|M_z(n, t_w \to \infty)| \leq \mathcal{A}^n. \tag{S17}$$

On the other hand, as established in Section V, the spin signal is a monotonically increasing function of $t_w$, meaning that the measured spin signal always satisfies $|M_z(n)| \leq \mathcal{A}^n$. Hence, the actual ensemble-averaged accuracy $\mathcal{A}$ is expected to be under-estimated by our fidelity measurement scheme.

## VII. DIPOLAR PERTURBATION METRIC CALCULATION

To incorporate the dipolar interactions in the optimization, the two-body perturbation Hamiltonian $\delta H = 2\sigma_z \otimes \sigma_z - \sigma_x \otimes \sigma_x - \sigma_y \otimes \sigma_y$ is used to minimize the first order correction to the final state $\mathcal{D}_{U \otimes U}(\delta H; T)|\psi_0\rangle^{\otimes 2}$, which is done by defining the perturbation metric as

$$\varphi_{\text{per}}(\mathbf{x}) \equiv 1 - \frac{1}{\mathcal{N}^2} \langle\psi_0|^{\otimes 2} \; \mathcal{D}_{U \otimes U}^\dagger(\delta H; T) \mathcal{D}_{U \otimes U}(\delta H; T) \; |\psi_0\rangle^{\otimes 2}, \tag{S18}$$

with $\mathcal{N} = \int_0^T dt \|\delta H\|_{\mathrm{op}} = 4T$ and $|\psi_0\rangle^{\otimes 2} = |\psi_0\rangle \otimes |\psi_0\rangle$. The metric is calculated by constructing the following Van Loan generator using a direct sum:

$$L(\mathbf{b}, t) \equiv \begin{bmatrix} -iH(\mathbf{b}) & P(\mathbf{b}) \\ 0 & -iH(\mathbf{b}) \end{bmatrix} \oplus \begin{bmatrix} -iH_2(\mathbf{b}) & -i\delta H \\ 0 & -iH_2(\mathbf{b}) \end{bmatrix} \tag{S19}$$

$$= -i \begin{bmatrix} H(\mathbf{b}) & iP(\mathbf{b}) & 0 & 0 \\ 0 & H(\mathbf{b}) & 0 & 0 \\ 0 & 0 & H_2(\mathbf{b}) & \delta H \\ 0 & 0 & 0 & H_2(\mathbf{b}) \end{bmatrix}, \tag{S20}$$

where $H_2(\mathbf{b}) \equiv H(\mathbf{b}) \otimes \mathbb{1} + \mathbb{1} \otimes H(\mathbf{b})$ is the two-spin unperturbed Hamiltonian. The associated Van Loan propagator is

$$V_t[\mathbf{b}] = \begin{bmatrix} U(\mathbf{x}, t) & \mathcal{D}_U(iP; t) & 0 & 0 \\ 0 & U(\mathbf{x}, t) & 0 & 0 \\ 0 & 0 & U(\mathbf{x}, t) \otimes U(\mathbf{x}, t) & \mathcal{D}_{U \otimes U}(\delta H; t) \\ 0 & 0 & 0 & U(\mathbf{x}, t) \otimes U(\mathbf{x}, t) \end{bmatrix}, \tag{S21}$$

as the time-ordered exponential respects the direct sum operation. With that in mind, the perturbation metric Eq.(S18) can be evaluated using the (3,4) block of $V_T[\mathbf{b}]$. The gradient calculation also follows the exact same steps, with the exception of the inverse $V_t^{-1}[\mathbf{b}]$, which can be determined analytically [8].

---


\end{document}